\documentclass{aa}
\usepackage{txfonts}
\usepackage{graphicx}
\begin{document}

\title{The long-period Galactic Cepheid RS\,Puppis}
\subtitle{I. A geometric distance from its light echoes\thanks{Based on observations collected at the European Southern Observatory, Chile under ESO Programme 078.D-0739(B).}}
\titlerunning{A geometric distance to RS\,Puppis from its light echoes}
\authorrunning{P. Kervella et al.}
\author{
P. Kervella\inst{1}
\and
A. M\'erand\inst{2}
\and
L. Szabados\inst{3}
\and
P. Fouqu\'e\inst{4}
\and
D. Bersier\inst{5}
\and
E. Pompei\inst{6}
\and
G. Perrin\inst{1}
}
\offprints{P. Kervella}
\mail{Pierre.Kervella@obspm.fr}
\institute{
LESIA, Observatoire de Paris, CNRS UMR 8109, UPMC, Universit\'e Paris Diderot, 5 Place Jules Janssen,
F-92195 Meudon, France
\and
Center for High Angular
Resolution Astronomy, Georgia State University, PO Box 3965,
Atlanta, Georgia 30302-3965, USA
\and
Konkoly Observatory, H-1525 Budapest XII, P. O. Box 67, Hungary
\and
Observatoire Midi-Pyr\'en\'ees, Laboratoire d'Astrophysique,
CNRS UMR 5572, Universit\'e Paul Sabatier - Toulouse 3,
14 avenue Edouard Belin, F-31400 Toulouse, France
\and
Astrophysics Research Institute, Liverpool John Moores University, Twelve Quays House, Egerton Wharf, Birkenhead, CH411LD, United Kingdom
\and
European Southern Observatory, Alonso de Cordova 3107,
Casilla 19001, Vitacura, Santiago 19, Chile
}
\date{Received ; Accepted}
\abstract
% Context
{The bright southern Cepheid RS\,Pup is surrounded by a circumstellar nebula reflecting the light from the central star. The propagation of the light variations from the Cepheid inside the dusty nebula creates spectacular light echoes that can be observed up to large distances from the star itself. This phenomenon is currently unique in this class of stars.}
% Aims
{For this relatively distant star, the trigonometric parallax available from Hipparcos has a low accuracy. A careful observation of the light echoes has the potential to provide a very accurate, geometric distance to RS\,Pup.}
% Methods
{We obtained a series of CCD images of RS\,Pup with the NTT/EMMI instrument covering the variation period of the star ($P=41.4$\,d), allowing us to track the progression of the light wavefronts over the nebular features surrounding the star. We measured precisely the phase lag of the photometric variation in several regions of the circumstellar nebula.}
% Results
{From our phase lag measurements, we derive a geometric distance of $1\,992 \pm 28$\,pc to RS\,Pup. This distance is affected by a total uncertainty of 1.4\%, and corresponds to a parallax of $\pi = 0.502 \pm 0.007$\,mas and a distance modulus of $\mu = 11.50 \pm 0.03$.}
% Conclusions
{The geometric distance we derive is by far the most accurate to a Cepheid, and among the most accurate to any star. RS\,Pup appears both as somewhat neglected and particularly promising to investigate the mass-loss history of Cepheids. Thanks to its highly accurate distance, it is also bound to become an important luminosity fiducial for the long period part of the period-luminosity diagram.}

\maketitle

%__________________________________Introduction
\section{Introduction}

The variability of \object{RS\,Pup} was discovered by Miss Reitsma, Kapteyn's assistant in 1897, and independently of her by Innes~(\cite{innes897}). The period of about 41.3 days, determined at the turn of the 20$^{\rm th}$ century, was the longest period among the Cepheids known at that time. Long period Cepheids are important because, owing to their larger mass, they evolve more rapidly than their shorter period counterparts. In addition, they are key objects in calibrating the period-luminosity relationship, and the exact knowledge of the behaviour of this relationship at the long-period end is critical for establishing the cosmic distance scale because being more luminous, long-period Cepheids can be discovered more easily in remote galaxies.

RS\,Puppis (\object{HD 68860}, \object{SAO 198944}) is one of the most important Cepheids in our Galaxy due to its peculiarities:\\
- it is embedded in a reflection nebula,\\
- it is situated in the region of a stellar association (Puppis\,III),\\
- its pulsation period, which is one of the longest known values among Galactic Cepheids, is subjected to strong variations.

In spite of the wealth of information that can be retrieved from thoroughly studying these peculiarities, RS\,Puppis has been a neglected Cepheid for 2-3 decades.

Before our interferometric works (Kervella et al.~\cite{kervella06}; M\'erand et al.~\cite{merand06}, \cite{merand07}), the nebula around \object{RS Pup} was the only certain case of a direct relationship between a Cepheid and interstellar material. Another Cepheid, \object{SU Cas}, also appears close to a reflection nebula (Milone et al.~\cite{milone99}; Turner \& Evans~\cite{turner84}), but the physical association is more uncertain.
The nebulosity around RS\,Pup was discovered by Westerlund~(\cite{westerlund61}). No other Cepheid is known to be so intimately associated with the interstellar matter. Westerlund already remarked that `the nebula may be large enough to permit the detection of possible variations in its intensity distribution due to the light variations of the star'. This offers a unique opportunity to determine the distance of the Cepheid by comparing the angular and linear radii of the denser rings in the reflection nebula. The positional coincidence of RS\,Pup with the reflection nebula can indeed be used for determining the distance and luminosity of the Cepheid, based on the light echo phenomenon occurring around such a large amplitude variable (Havlen~\cite{havlen72a}; see also Sugerman~\cite{sugerman03}). This distance is independent of the results obtained by other methods applicable for such variable stars. Another aspect of this positional coincidence is the study of the relationship between stellar pulsation and matter expelled from the star.

Moreover, Westerlund~(\cite{westerlund63}) pointed out that the reflection nebula surrounding RS\,Pup is a part of a stellar association consisting of early-type stars, which allowed him to derive the distance of the newly detected Pup\,III association. The spatial coincidence of the Pup\,III association and RS\,Pup allows another determination of the distance of the Cepheid independently of the methods based on the pulsation. 

Ring-like structures in the nebula surrounding the Cepheid imply that the circumstellar matter, at least partly, originates from the pulsating atmosphere of RS\,Pup. The mechanism for the formation of this circumstellar envelope (CSE) is currently unknown. Deasy~(\cite{deasy88}) proposed several scenarios, ranging from evolution of the star before its Cepheid phase to mass loss during multiple crossings of the instability strip. As pointed out by Havlen~(\cite{havlen72a}) and Szabados~(\cite{szabados03}), it is unlikely that RS Pup is the only existing Cepheid--nebula association. Based on IRAS photometry and IUE ultraviolet spectra, Deasy~(\cite{deasy88}) identified mass loss in a number of Cepheids. The highest mass loss rate is attributed by this author to RS\,Pup ($10^{-6}\,M_{\odot}.{\rm yr}^{-1}$). This level is very significant, comparable to that of the Mira stars (which host dense molecular and dusty envelopes), and explains the ``bubble" structures that have been carved by this star in the interstellar medium.

The available observational material on RS\,Pup mostly consists of photometric data covering more than a century. The analysis of these data is instrumental in determining the history of variations in the pulsation period, which will be discussed in a forthcoming paper. Changes in the pulsation period bear information on the evolutionary stage of the Cepheid (secular period changes) and the physical processes occurring in the pulsating stellar atmosphere (period changes on time-scale of years). Recent high resolution spectroscopic observations with the HARPS instrument were reported by Nardetto et al.~(\cite{nardetto06}, \cite{nardetto07}). They show very broad spectral lines at certain phases, indicative of the presence of strong shock waves.

In the present article, we describe our application of the light echo method to determine the distance to RS\,Pup. The observations obtained with ESO's NTT are reported in Sect.~\ref{observations}. A detailed description of the processing steps that were applied to derive the photometry of the nebula is given in Sect.~\ref{data-analysis}. Finally, we derive the distance to RS\,Pup in Sect.~\ref{distance}, and compare our value to existing estimates in Sect.~\ref{discussion}.

%__________________________________Observations
\section{Observations \label{observations}}

We observed RS\,Pup using the 3.6\,m ESO New Technology Telescope (NTT) installed at La Silla Observatory (Chile), and equipped with the ESO Multi-Mode Instrument (EMMI).  This instrument is a multipurpose imager and spectrograph (Dekker et al.~\cite{dekker86}). For the reported observations, we used the red arm of this instrument, based on two back-illuminated 
2k$\times$4k CCDs (15\,$\mu$m pixels) arranged in a mosaic, with a 7.82\,arcsec gap between the two detectors. The useful field of view is 9.0$\times$9.9\,arcmin. We restricted our processing to 5.0$\times$5.0 arcmin, still leaving a considerable margin around the nebula of RS\,Pup. In order to avoid a heavy saturation of the detector, we positioned RS\,Pup in the gap between the detectors. The Nasmyth adapter of the telescope was rotated in order to align the direction of the gap with a lower density part of the nebula. The pixel scale of EMMI is $0.1665 \pm 0.0006^{\prime\prime}$.pixel$^{-1}$ on the sky\footnote{http://www.ls.eso.org/lasilla/sciops/ntt/emmi/newRed/}, providing a good sampling of the point spread function (PSF) even under good seeing conditions.

%__________________________________Table of observations
\begin{table*}
\caption{Log of the EMMI observations of RS\,Pup and its PSF calibrator, HD\,70555. MJD$^*$ is the median Modified Julian Date of the exposures. The phases $\phi$ were computed using the $(T_0, P)$ parameters given in Sect.~\ref{photometryRSPup}.}
\label{emmi_log}
\begin{small}
\begin{tabular}{ccccccccc}
\hline
Date & UTC & MJD$^*$ & Star & Filter & $\phi$ & Exposures & Airmass & FWHM\,($^{\prime\prime}$) \\
\hline
2006-10-28 & 8:12:23 & 54036.3419 & HD\,70555 & $B$ &  & 25\,s $\times$ 6 & 1.17 & 1.19 \\
2006-10-28 & 8:37:14 & 54036.3592 & RS\,Pup & $B$ & 0.6854 & 180\,s $\times$ 10 & 1.10 & 1.14 \\
2006-10-29 & 7:54:05 & 54037.3292 & HD\,70555 & $B$ &  & 25\,s $\times$ 6 & 1.21 & 1.25 \\
2006-10-29 & 8:23:17 & 54037.3495 & RS\,Pup & $B$ & 0.7093 & 180\,s $\times$ 10 & 1.12 & 1.13 \\
2006-12-20 & 3:30:42 & 54089.1463 & RS\,Pup & $B$ & 0.9592 & 180\,s $\times$ 10 & 1.40 & 2.01 \\
2006-12-20 & 3:57:31 & 54089.1649 & HD\,70555 & $B$ &  & 25\,s $\times$ 6 & 1.33 & 1.97 \\
2007-01-13 & 5:28:49 & 54113.2283 & RS\,Pup & $B$ & 0.5404 & 180\,s $\times$ 10 & 1.01 & 0.91 \\
2007-01-13 & 5:54:39 & 54113.2463 & HD\,70555 & $B$ &  & 25\,s $\times$ 6 & 1.01 & 0.70 \\
2007-01-22 & 2:39:17 & 54122.1106 & HD\,70555 & $B$ &  & 25\,s $\times$ 7 & 1.16 & 1.16 \\
2007-01-22 & 3:09:38 & 54122.1317 & RS\,Pup & $B$ & 0.7552 & 180\,s $\times$ 12 & 1.08 & 0.96 \\
2007-01-27 & 3:49:50 & 54127.1596 & HD\,70555 & $B$ &  & 25\,s $\times$ 6 & 1.02 & 0.88 \\
2007-01-27 & 3:55:38 & 54127.1636 & HD\,70555 & $V$ &  & 6\,s $\times$ 6 & 1.02 & 1.13 \\
2007-01-27 & 4:00:11 & 54127.1668 & HD\,70555 & $R$ &  & 2\,s $\times$ 6 & 1.01 & 0.98 \\
2007-01-27 & 4:04:58 & 54127.1701 & HD\,70555 & $I$ &  & 3\,s $\times$ 6 & 1.01 & 0.98 \\
2007-01-27 & 4:18:53 & 54127.1798 & RS\,Pup & $B$ & 0.8770 & 179\,s $\times$ 5 & 1.01 & 0.88 \\
2007-01-27 & 4:35:02 & 54127.1910 & RS\,Pup & $V$ & 0.8773 & 45\,s $\times$ 8 & 1.00 & 0.89 \\
2007-01-27 & 4:47:06 & 54127.1994 & RS\,Pup & $R$ & 0.8775 & 20\,s $\times$ 12 & 1.01 & 0.80 \\
2007-01-27 & 4:59:11 & 54127.2078 & RS\,Pup & $I$ & 0.8777 & 30\,s $\times$ 10 & 1.01 & 0.81 \\
2007-03-24 & 0:16:25 & 54183.0114 & RS\,Pup & $B$ & 0.2244 & 180\,s $\times$ 10 & 1.01 & 0.88 \\
2007-03-24 & 0:44:32 & 54183.0309 & HD\,70555 & $B$ &  & 25\,s $\times$ 6 & 1.00 & 0.74 \\
\hline
\end{tabular}
\end{small}
\end{table*}

Immediately before or after the RS\,Pup observations, we observed as a PSF reference the K1III star \object{HD\,70555}. We observed this star in addition to RS\,Pup in order to subtract the wings of the diffraction pattern of RS\,Pup from the nebula.
We obtained a series of exposures through the $B$ filter of EMMI at seven epochs, complemented by a single epoch of $V$, $R$, and $I$\footnote{The transmission curves of the EMMI filters are available from http://www.ls.eso.org/lasilla/sciops/ntt/emmi/emmiFilters.html}.
%The central wavelength of the $B$ filter is 413.9\,nm, with a FWHM of 109.8\,nm.
The images in $VRI$ colors will be discussed in a forthcoming article. The exposures were obtained randomly between 28 October 2006 and 24 March 2007, but resulted in a reasonably good coverage of the pulsation cycle of the star. 
The individual exposure times were relatively short, between 20 and 180\,s, in order to avoid the saturation of the detector even very close to the star. The journal of the
observations is presented in Table~\ref{emmi_log}. The total shutter open
time obtained on RS\,Pup with EMMI in all bands reaches more than 3.5\,hours.
The raw frames produced by the instrument were pre-processed in a standard way (bias removal, flat-fielding, cropping) using the Yorick\footnote{http://yorick.sourceforge.net/} and IRAF\footnote{IRAF is distributed by the National Optical Astronomy Observatories, which are operated by the Association of Universities for Research in Astronomy, Inc., under cooperative agreement with the National Science Foundation.} software packages. The resulting frames were precisely co-aligned using the IRAF script {\tt imcoadd} ({\tt gemini} package).

In order to better vizualize the light echo phenomenon, we computed a continuous movie of the progression of the light echoes in the nebula, using a spline interpolation of the measured epochs for each pixel (excepting the night of 20 December 2006). Snapshots of this interpolated movie are presented in Fig.~\ref{movie} for ten regularly spaced phases, and the movie itself (50 interpolated epochs) is available from the A\&A web site in MPEG format. The variation of the luminosity of RS\,Pup is visible on these images from the changing extension of the wings of its PSF. The morphological changes of the nebula with time are considerable, and they will be further studied in a forthcoming paper. This movie was not used for the analysis developed below.

%______________ Figure
\onlfig{1}{
\begin{figure*}[ht]
\centering
\includegraphics[width=6cm, angle=0]{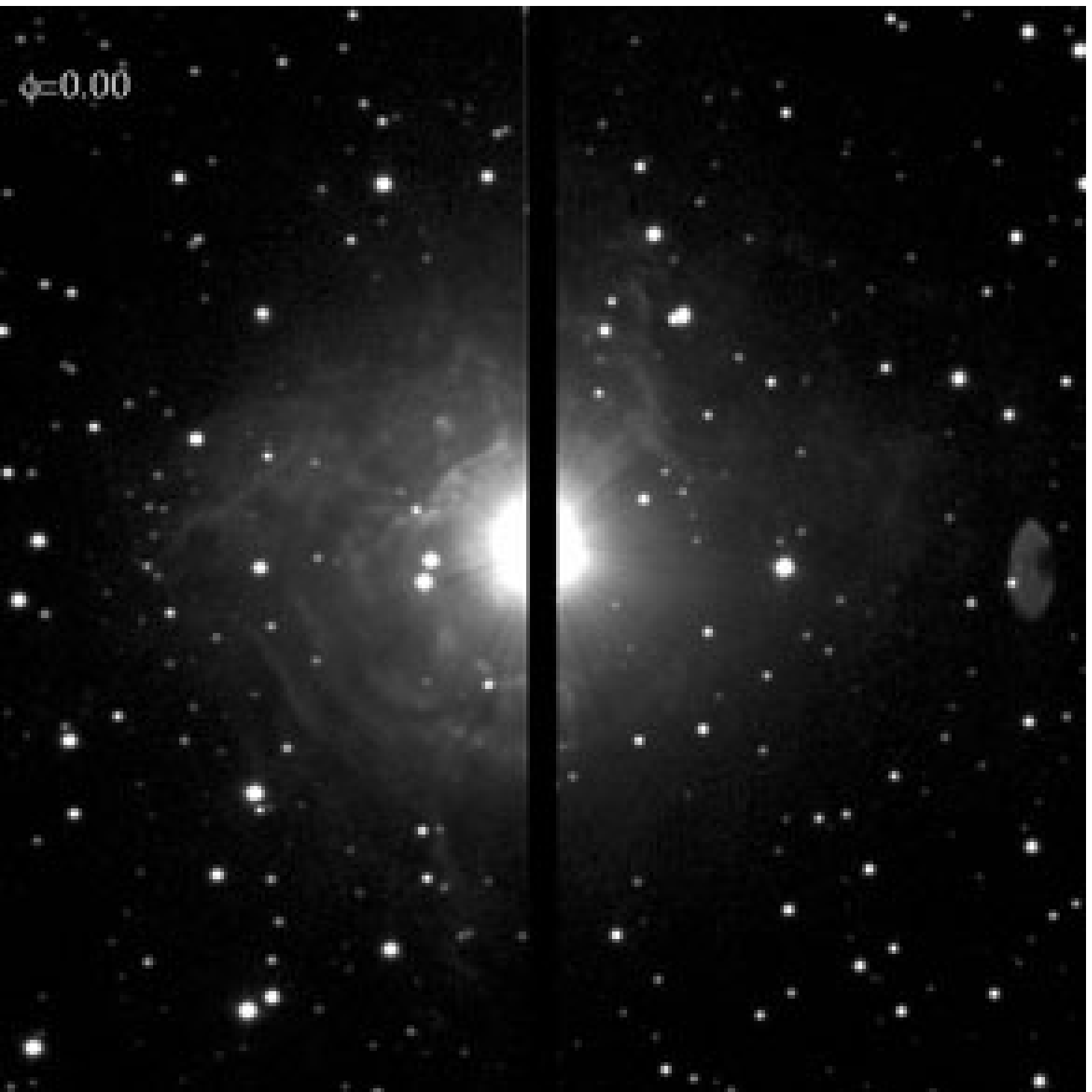}
\includegraphics[width=6cm, angle=0]{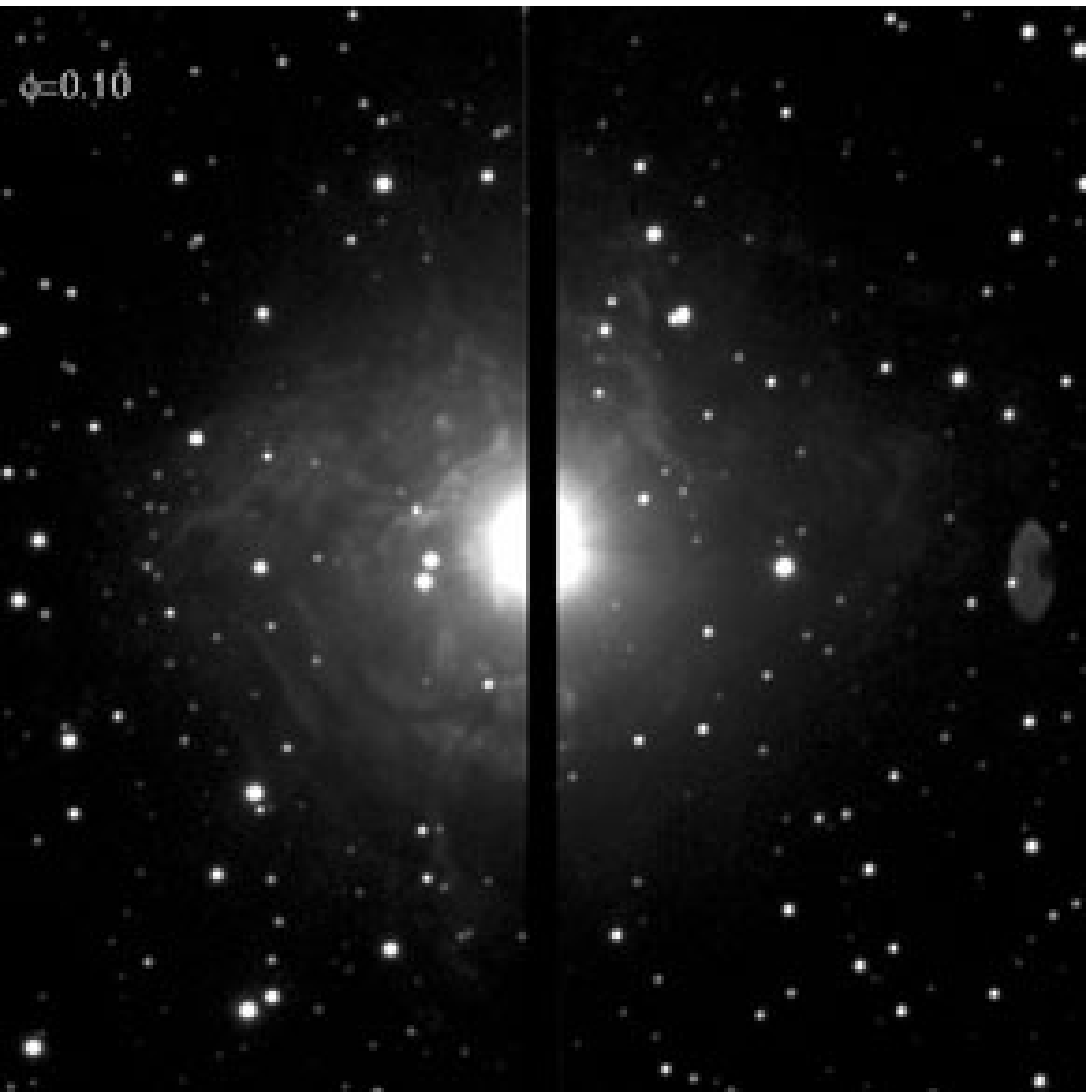}
\includegraphics[width=6cm, angle=0]{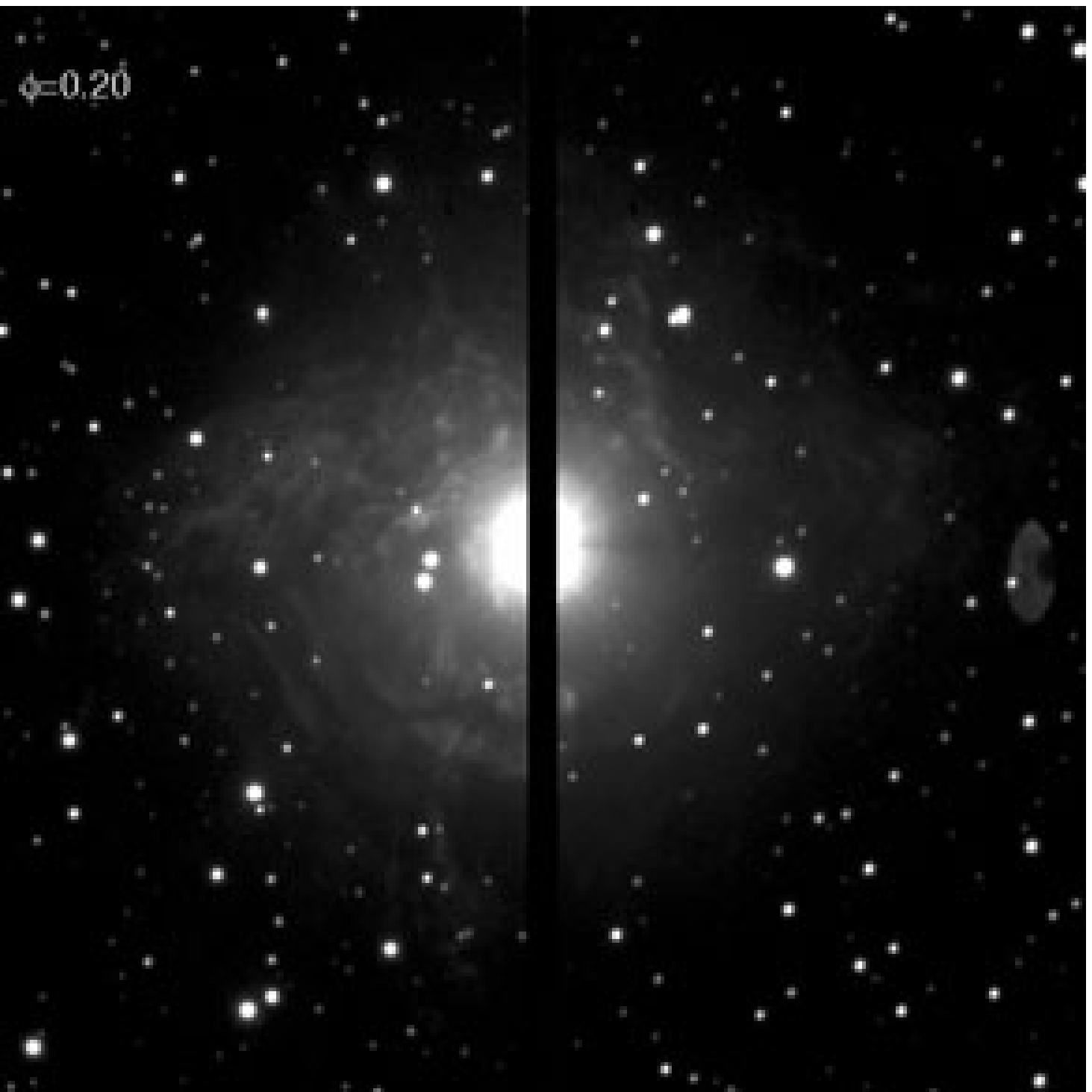}
\includegraphics[width=6cm, angle=0]{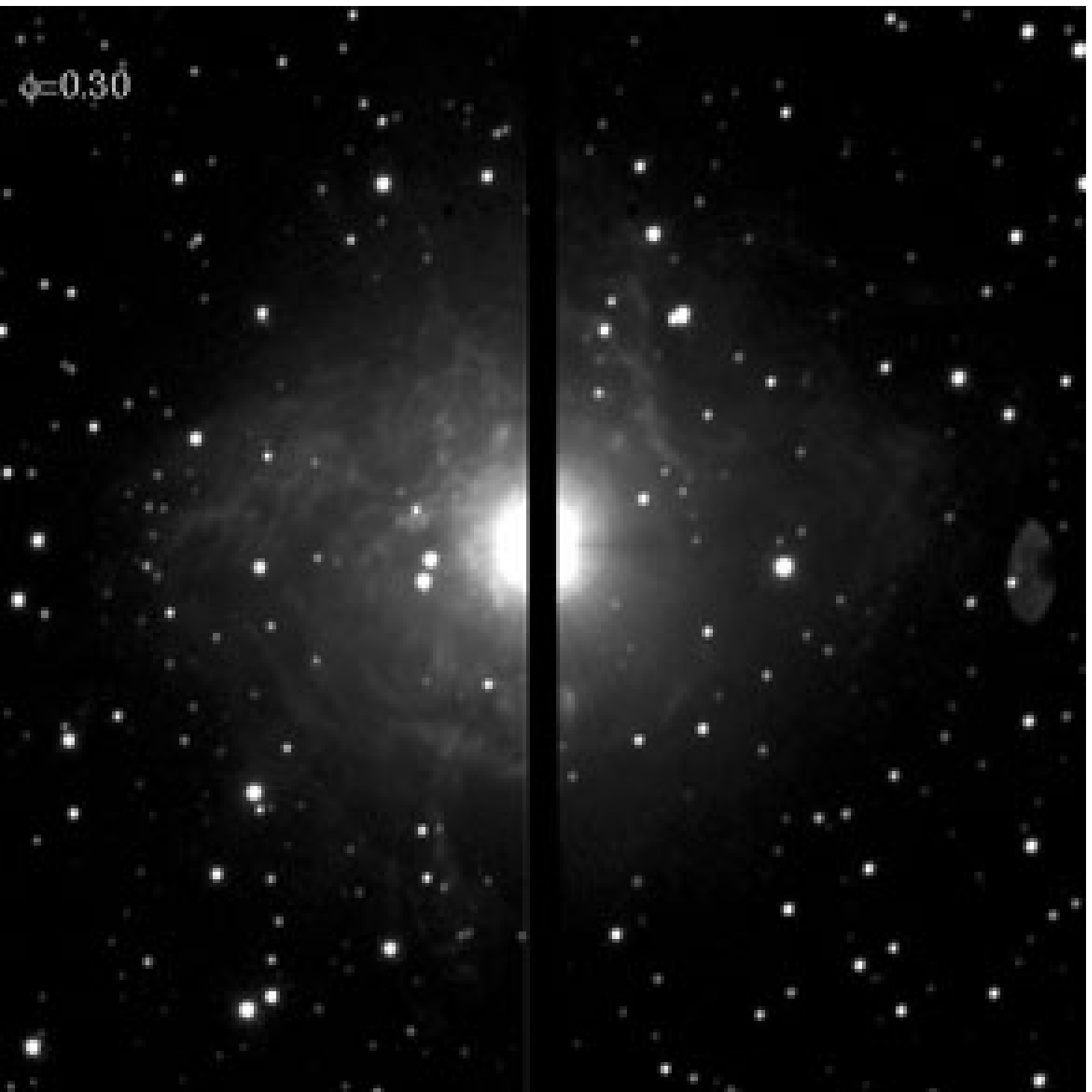}
\includegraphics[width=6cm, angle=0]{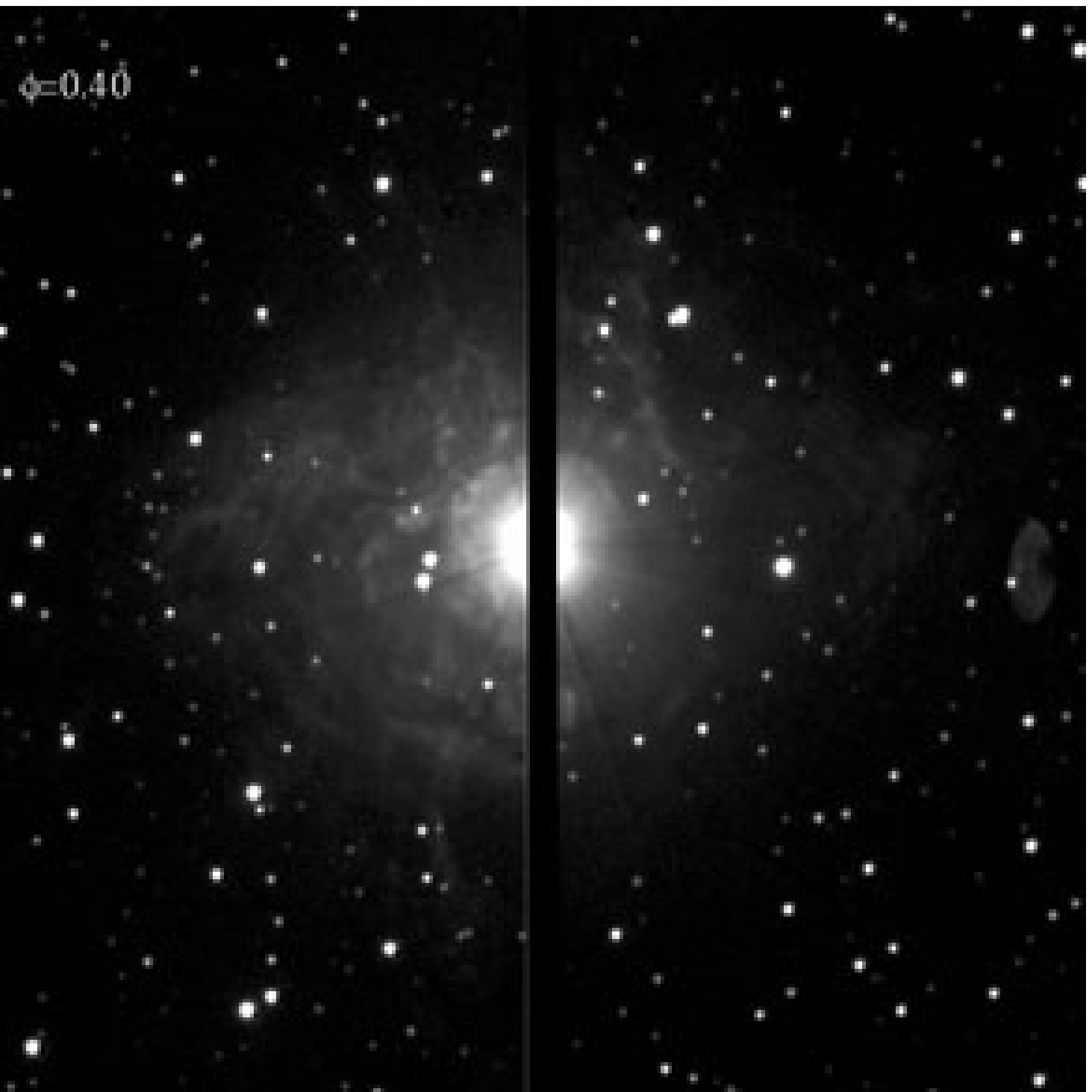}
\includegraphics[width=6cm, angle=0]{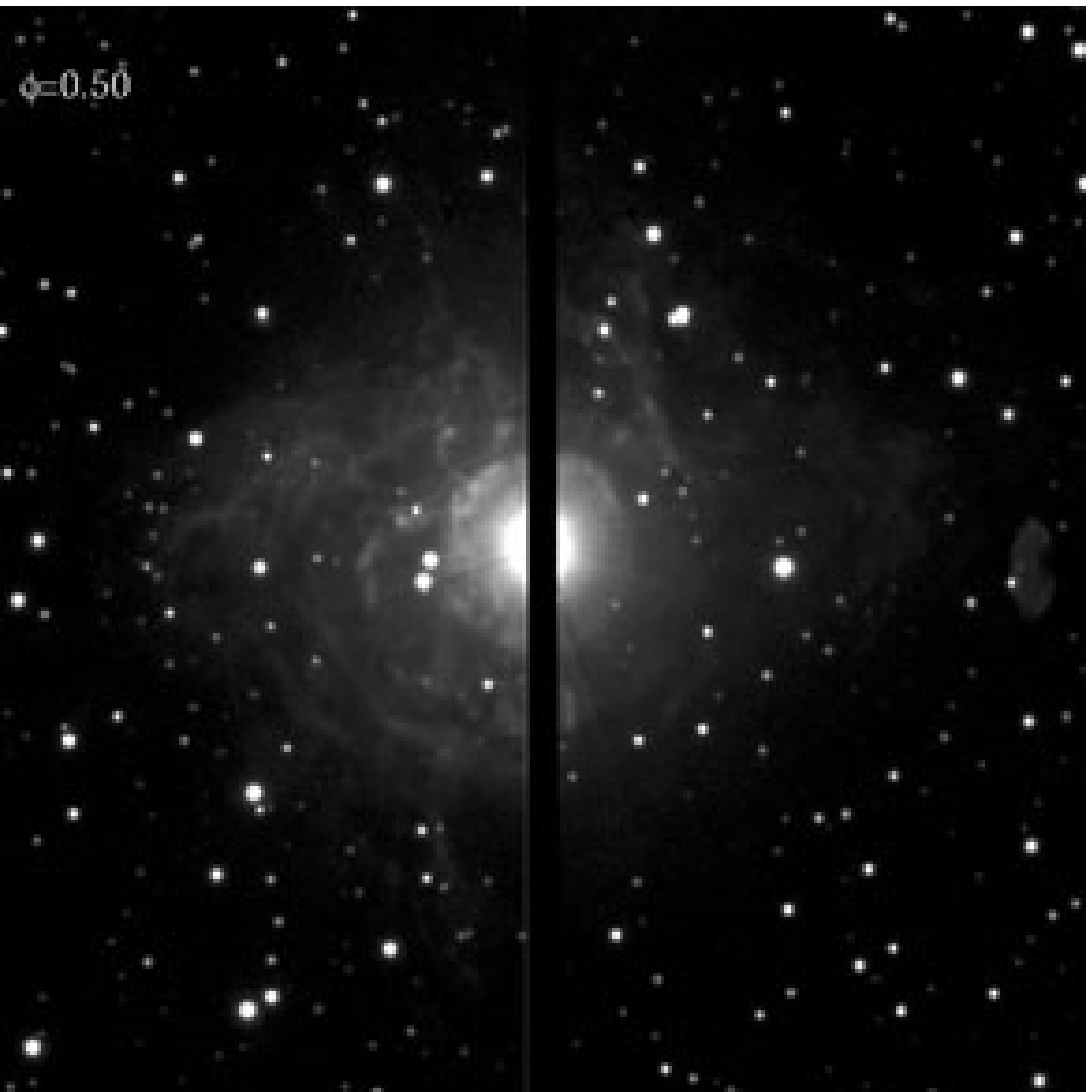}
\includegraphics[width=6cm, angle=0]{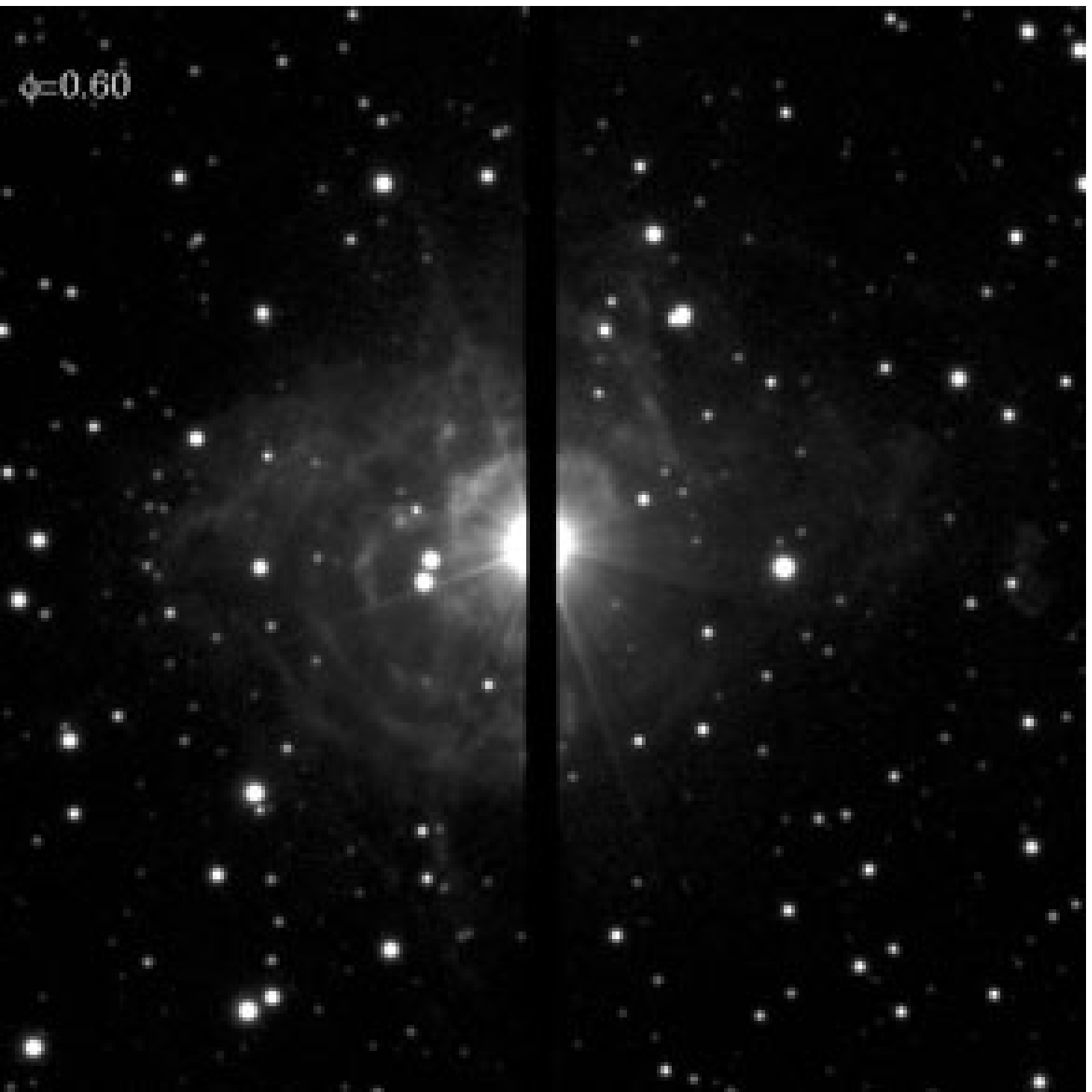}
\includegraphics[width=6cm, angle=0]{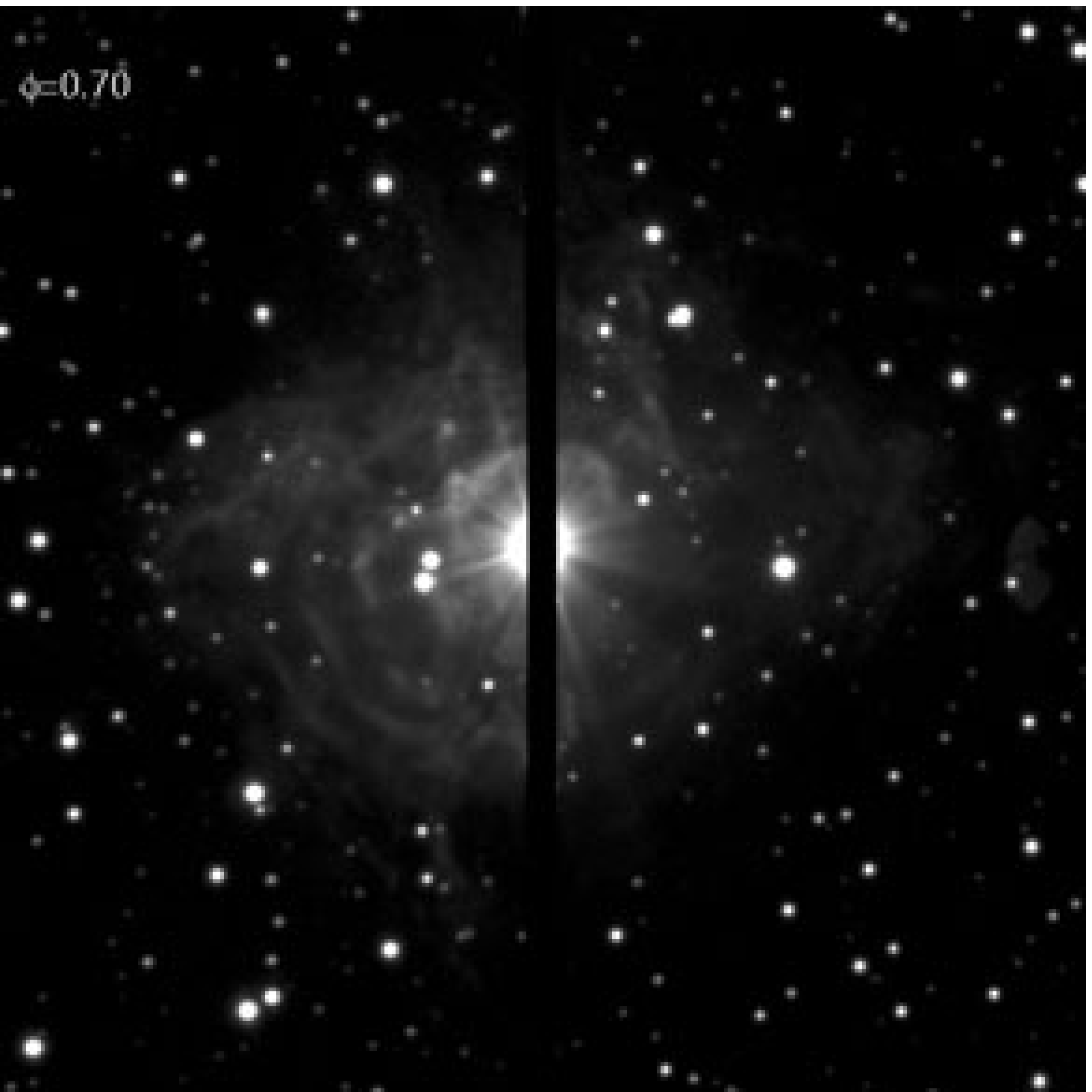}
\includegraphics[width=6cm, angle=0]{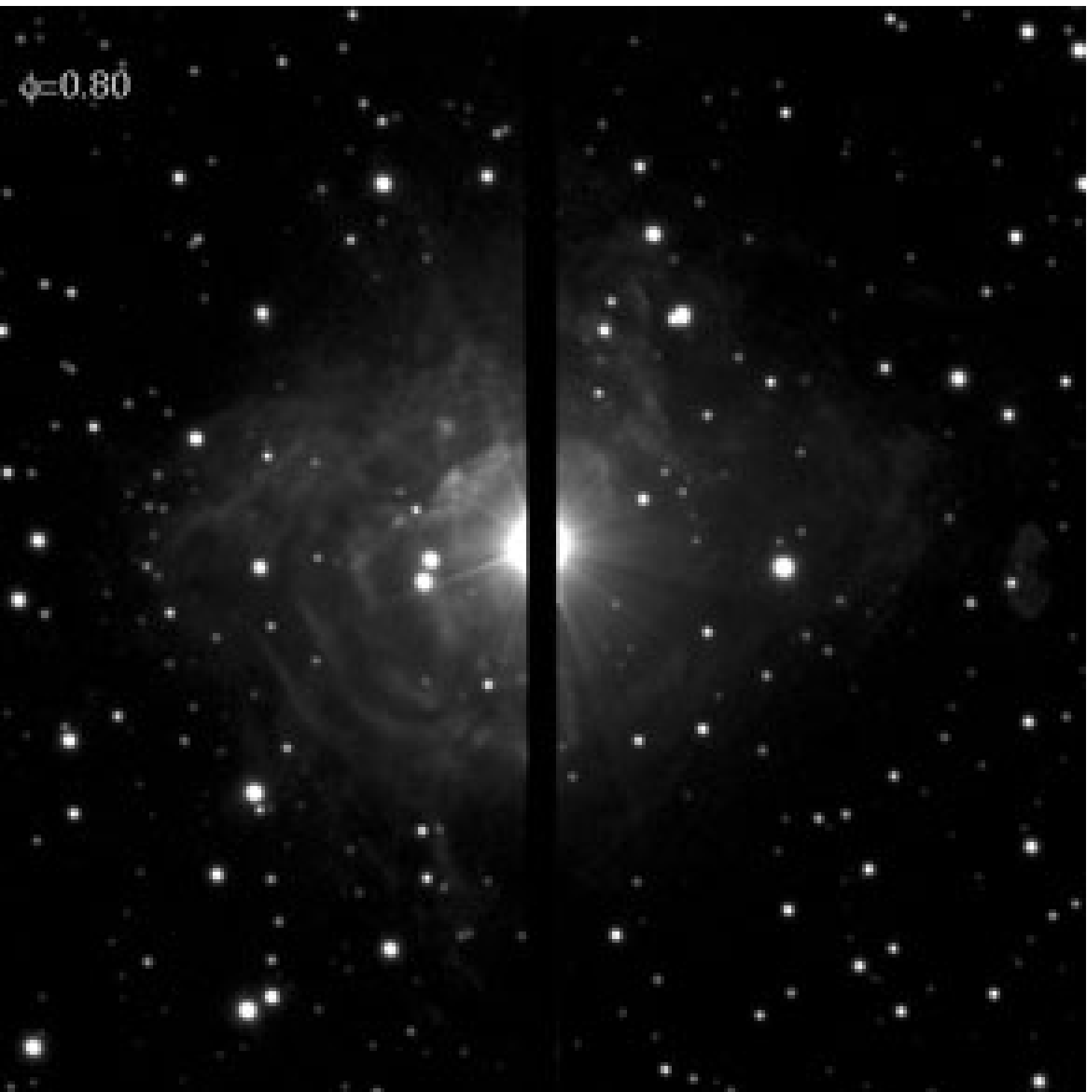}
\includegraphics[width=6cm, angle=0]{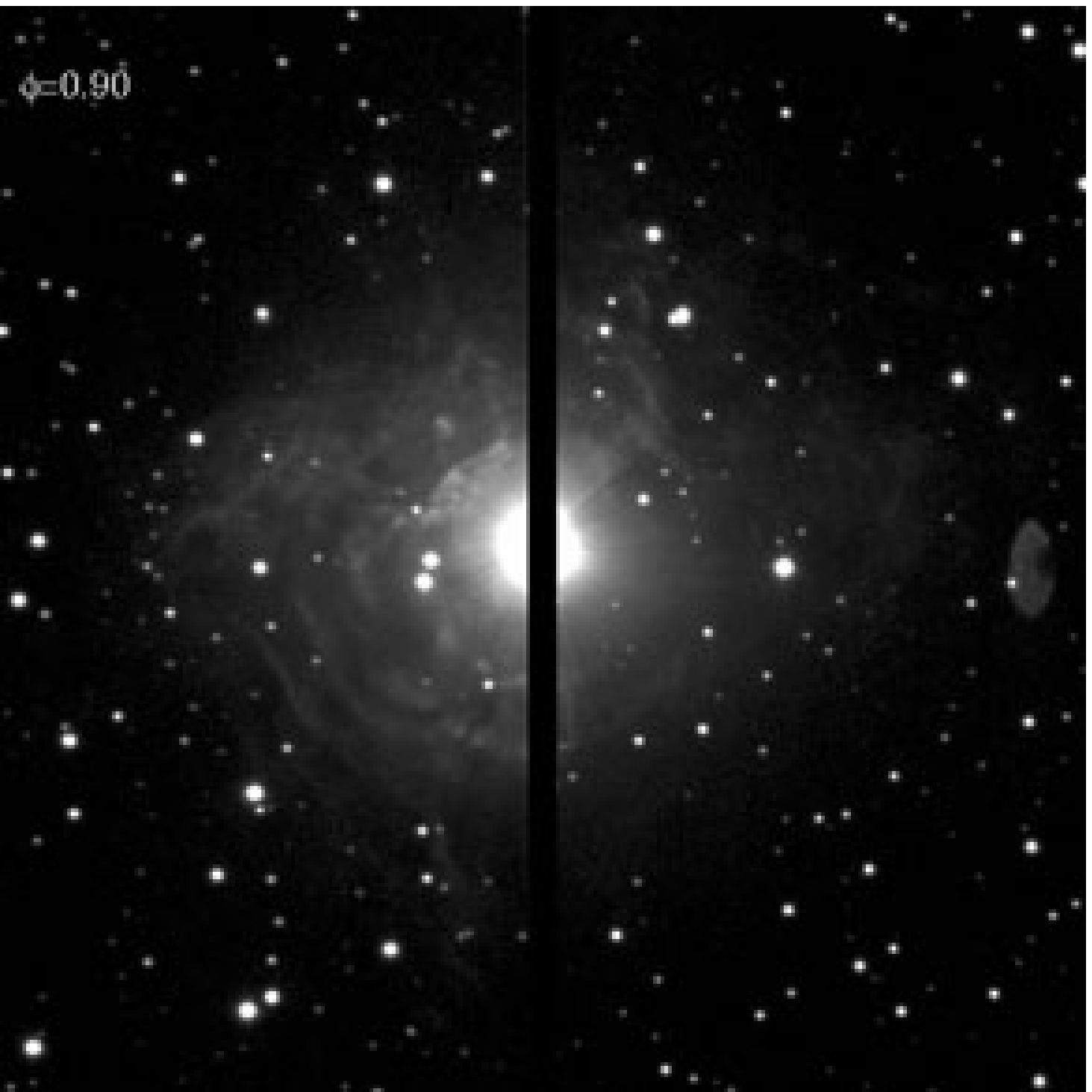}
\caption{Snapshots of a spline interpolation of our observations of RS\,Pup. An MPEG movie showing the progression of the wavefronts in the nebula is available through the A\&A web site. A ghost image of part of the telescope pupil can be seen on the right of the images.}
\label{movie}
\end{figure*}
}

To obtain absolute coordinates on a common world coordinate system (WCS), we retrieved from the IPAC archive\footnote{http://irsa.ipac.caltech.edu/} the FITS images covering the field of view of our images. The astrometric calibration was obtained by matching the coordinates of the brightest sources with their counterparts in the 2MASS archive images. The absolute astrometric accuracy of the 2MASS images is about 0.1$^{\prime\prime}$
(Skrutskie et al.~\cite{skrutskie06}). Due to the presence of stronger scattered light in the visible, our astrometric uncertainty is estimated to be of the order of 0.2$^{\prime\prime}$ ($\approx$ one\,pixel) over the EMMI field. As we do not aim at obtaining absolute astrometry, this calibration method is sufficient.

%__________________________________Analysis
\section{Data analysis \label{data-analysis}}

\subsection{PSF subtraction and photometry \label{psf-sub}}

\subsubsection{Preparation of the reference PSF \label{prep-psf}}

\begin{figure}[ht]
\centering
\includegraphics[width=9.0cm, angle=0]{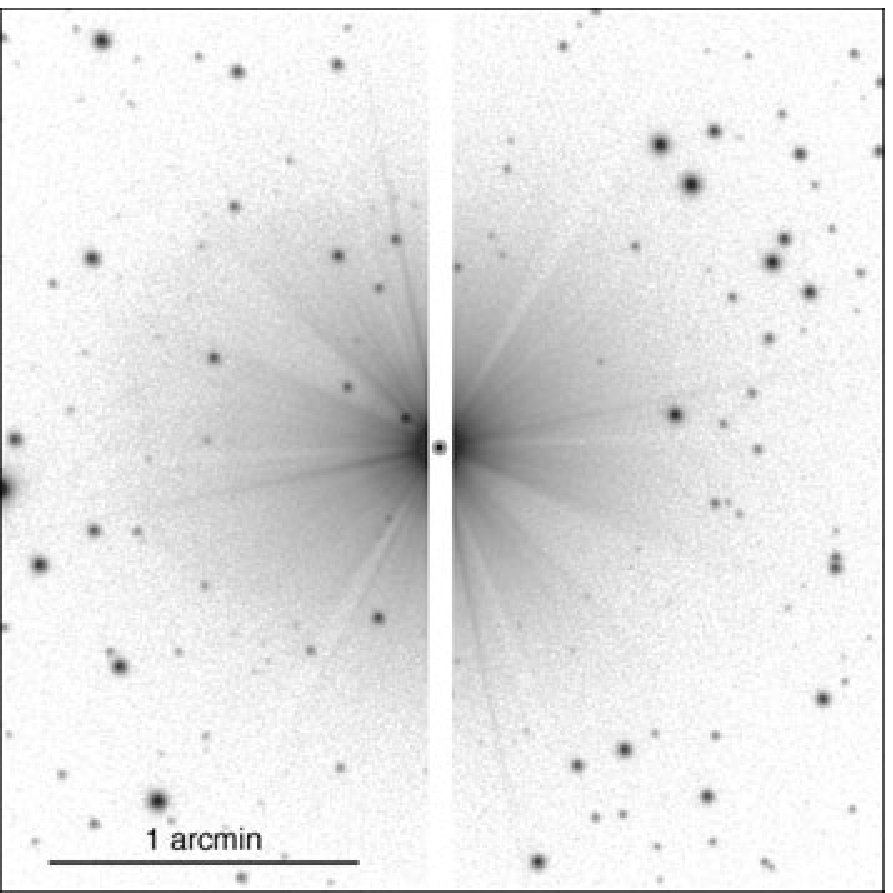}

\vspace{0.2cm}

\includegraphics[width=9.0cm, angle=0]{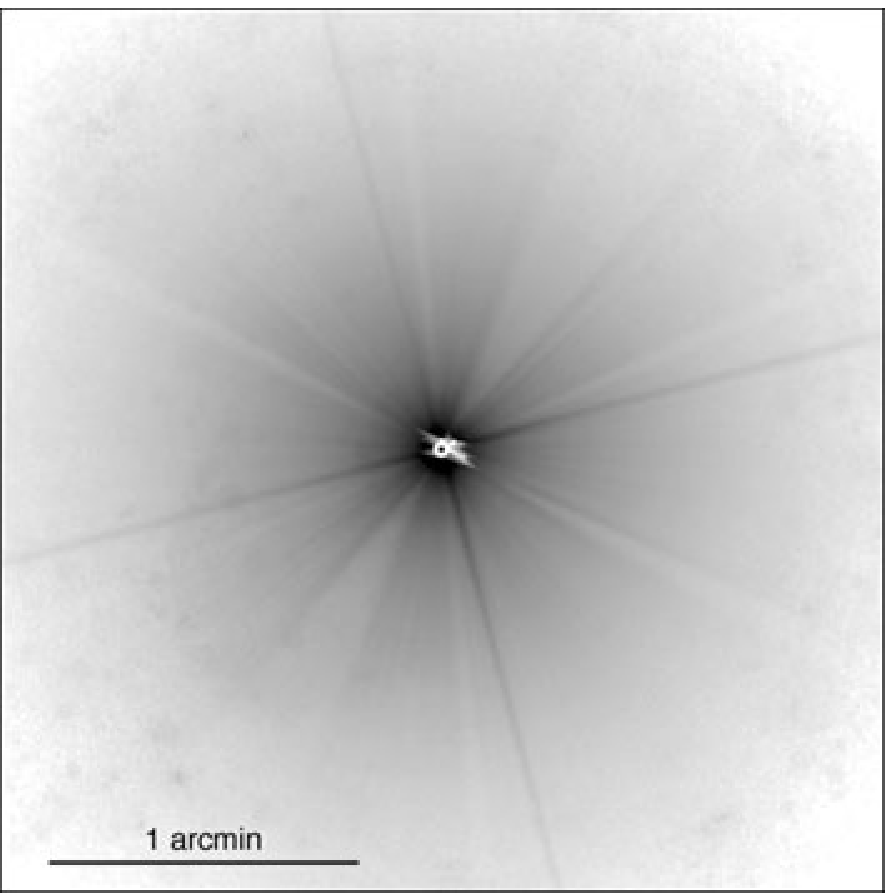}
\caption{{\it Top:} Image of the field around HD\,70555 showing the wings of the PSF of the NTT and the background stars. {\it Bottom:} Reference NTT/EMMI PSF extracted from the observations of HD\,70555 obtained over four epochs.}
\label{HD70555psf}
\end{figure}

On the images of RS\,Pup, large spikes are created by the secondary mirror spider support and the inhomogeneities of the optical surfaces in the telescope. Fortunately, from one star to the other and over long periods of time, the distribution of light in this diffraction pattern is very much reproducible, and it can be efficiently measured and subtracted.
We therefore observed immediately before or after RS\,Pup the bright K1III star \object{HD\,70555}. As shown in Fig.~\ref{HD70555psf} (top), the field around this star is relatively rich in background stars, due to its low Galactic latitude (+2$^\circ$). Its position near RS\,Pup is an advantage to match the airmasses, but the presence of background stars, which must be removed before subtraction from the RS\,Pup images, creates some difficulties.

Due to the alt-az configuration of the NTT, the pupil of the telescope (and all the attached PSF defects) rotate at an angular rate that is a function of the altitude and azimuth of the object. The position angle of the diffraction spikes is given by the parallactic angle at the time of the exposure.
These optical defects are caused by the secondary mirror support and the optical elements, and therefore they rotate with the telescope pupil. During normal operation of the telescope, the field de-rotator (a mechanical device on which the instrument is mounted) takes care of rotating continuously the instrument in order to keep the position of the objects constant on the detector. Of course, any defect introduced by the instrument itself does not rotate with the telescope pupil (such as ghosts created by internal reflections, e.g. inside filters).

To take advantage of this effect, we ``re-rotated" the PSF exposures obtained on HD\,70555 for the nights of 2006-10-28, 2007-01-13, 2007-01-27 and 2007-03-24. In other words, using the parallactic angle information, we aligned precisely the PSF spikes with each other, letting the background star field rotate.
We chose these four nights as their seeing FWHM is comparable (from 0.70$^{\prime\prime}$ to 1.14$^{\prime\prime}$), and they are all different in terms of parallactic angle. The effective FWHMs of these exposures were then all equalized to 1.19$^{\prime\prime}$, using the {\tt psfmatch} task of IRAF.
This command convolves the images by a kernel that is computed to match precisely the images to a reference frame (in this case, the first observation of HD\,70555 obtained under slightly degraded seeing conditions on the night of 2006-10-28). Computing the median of these re-rotated images resulted in a very efficient removal of the background objects, leaving only the PSF of the central star, as shown in Fig.~\ref{HD70555psf} (bottom).
The latter image is rotated to a null parallactic angle, and the image scale is logarithmic to show the faint extensions of the PSF wings. The background stars disappeared almost completely. As in all the images presented in this article, the position of the observed star is marked with a spot.
Apart from the small artefacts close to the center, this image shows the actual PSF produced by the NTT with a seeing of 1.19$^{\prime\prime}$, for a null parallactic angle, i.e., with the altitude direction of the telescope aligned along the vertical axis of the image.
We finally prepared an adapted reference PSF frame for each individual image of RS\,Pup by re-rotating the image presented in Fig.~\ref{HD70555psf} (bottom) to the same parallactic angle.

%______________ Figure

\subsubsection{Preparation of the images of RS\,Pup \label{rspup_prep}}

In order to properly subtract the image obtained at Sect.~\ref{prep-psf}, we matched the PSFs of each RS\,Pup image to the same FWHM of 1.19$^{\prime\prime}$, which is also the FWHM of the PSF reference image. We normalized photometrically all images to the first one using a sample of twenty field stars and aperture photometry (with a large 20\,pixels aperture) with the {\tt phot} task of IRAF. The measured standard deviation of the photometry obtained on these twenty stars using a 5 pixels aperture is $\pm 0.03$\,mag over our complete image sample.

The night of 20 December 2006 is a special case as the very large FWHM is not caused by a poor seeing, but by a defocus of the telescope (the true seeing was around 0.7$^{\prime\prime}$). In order to recover useful information from this night, we used a special processing method. For this particular night, we degraded the FWHM of the reference PSF image from Sect.~\ref{prep-psf} to match the 2.01$^{\prime\prime}$ FWHM of the RS\,Pup image. We thus obtained a special PSF image set matching this series of images.

\subsection{PSF subtraction and photometry of RS\,Pup \label{photometryRSPup}}

%______________ Figure
\begin{figure}[ht]
\centering
\includegraphics[width=9.0cm, angle=0]{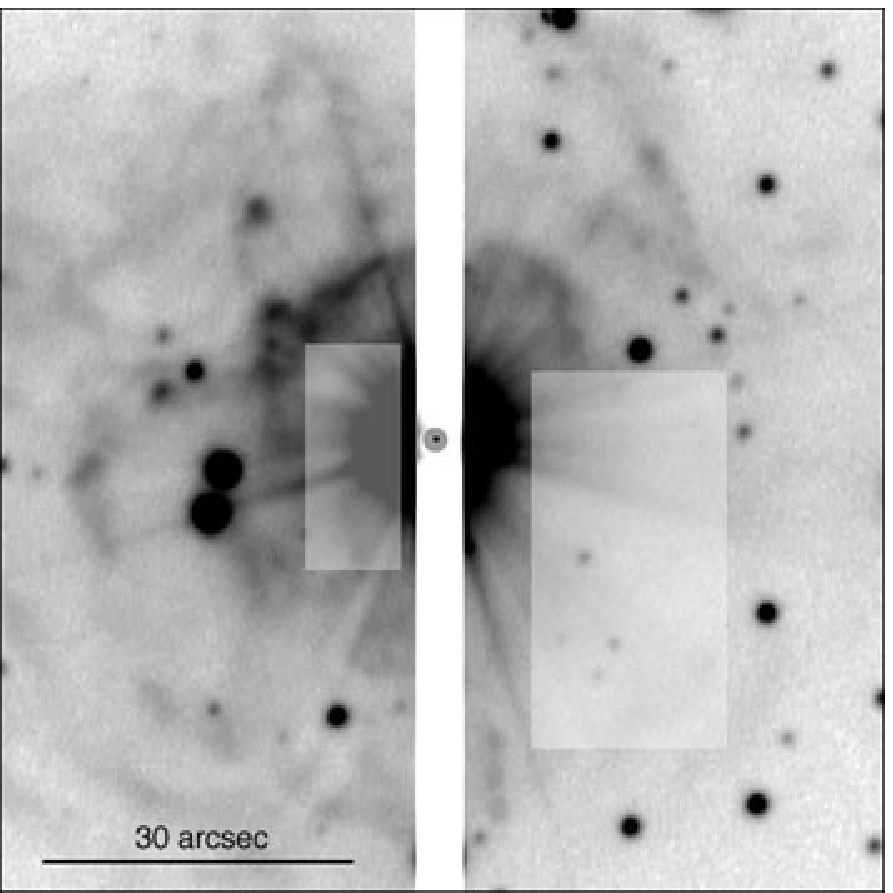}

\vspace{0.2cm}

\includegraphics[width=9.0cm, angle=0]{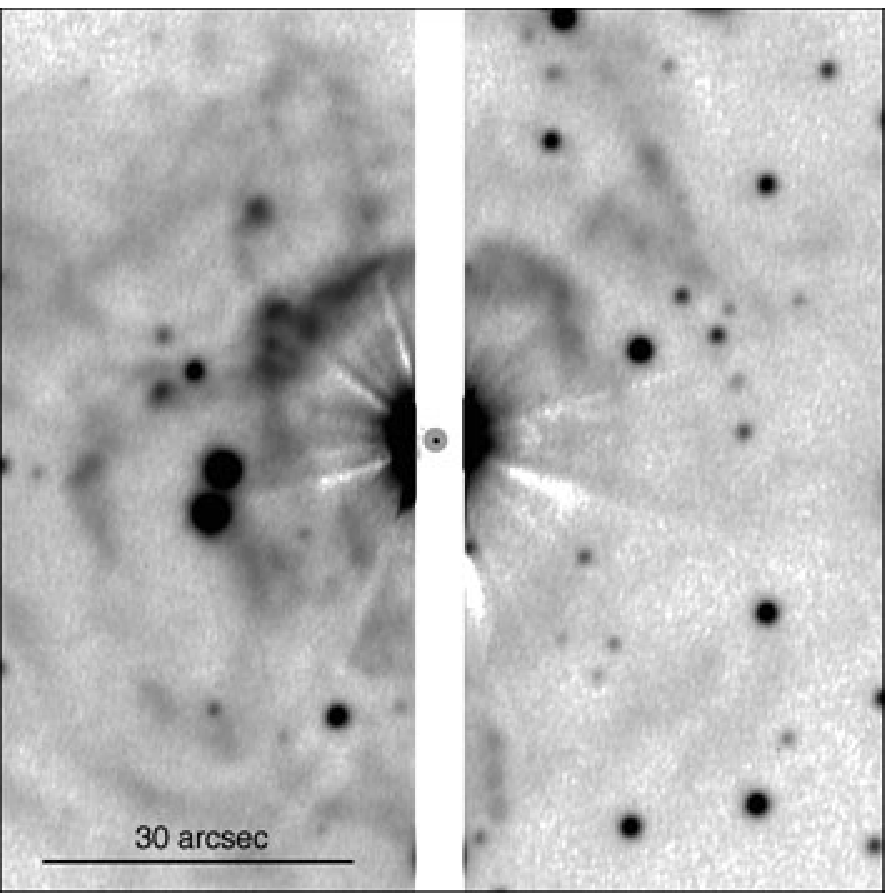}

\caption{{\it Top:} Rectangular zones (shown as light grey patches) selected to compute the photometric ratio between RS\,Pup and the reference PSF frame, superimposed on a single frame obtained on 28 October 2006. {\it Bottom:} Result of the PSF subtraction of the top frame.}
\label{RSPupzones}
\end{figure}

The goal of the whole procedure described in this section is to subtract the wings of the RS\,Pup PSF without affecting the nebular features over which the light echoes propagate. For this purpose, it is necessary to compute the flux ratio between RS\,Pup and HD\,70555. We initially thought about computing this ratio from predicted magnitudes of RS\,Pup, but in practice this is not possible due to its irregularly evolving period. This makes it difficult to phase precisely our measurements over archive photometry (even if only a few years old). So we resorted to a direct measurement of the flux in the wings of the PSF of RS\,Pup. However, the measurement itself is made difficult by the presence of the relatively bright nebular features that tend to bias the photometry towards higher flux.

\begin{figure}[ht]
\centering
\includegraphics[width=9cm, angle=0]{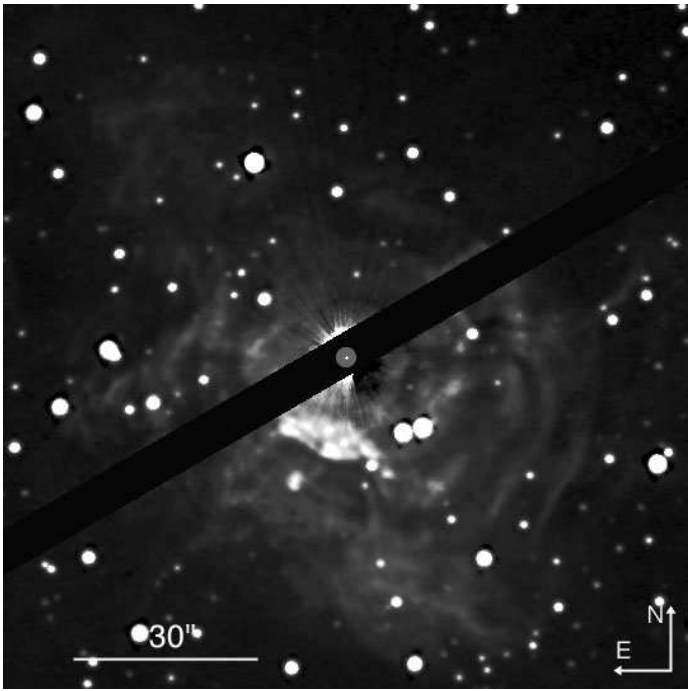}
\caption{PSF-subtracted image of the RS\,Pup nebula obtained on the night of 27 January 2007 at a phase $\phi = 0.8770$.}
\label{image-sample}
\end{figure}

We thus selected two regions close to RS\,Pup that show a low nebular contribution on all images (Fig.~\ref{RSPupzones}, top). Over these two regions, we smoothed the result of the division of each image of RS\,Pup by the corresponding PSF image, using a median box filter. We then took the minimum value as the ratio between the two frames. This method ensures that we do not overestimate the flux from RS\,Pup due to the nebula. As an example, the result of the subtraction is shown in Fig.~\ref{RSPupzones} (bottom) for a single frame. For each night, we computed the median of all subtracted images, an operation that further reduced the residuals of the PSF subtraction.
As an example of the result of this processing, the image of RS\,Pup obtained on 27 January 2007 is shown in Fig.~\ref{image-sample}. As visible in this image, the field de-rotator of the NTT was offset by 60$^\circ$ for the observations to avoid masking dense parts of the nebula inside the CCD gap. North is up and east is to the left on this image, but the other figures of this article are not corrected for the rotational offset. 

%______________ Figure

The derived flux ratio is the same between the frames obtained over one night to a few percent. We used the standard deviation of this dispersion as the statistical uncertainty of the ratio. To estimate the systematic error, we obtained aperture photometry of a sample of faint stars close to RS Pup (but unaffected by the nebula), in the subtracted images. The dispersion of the measured magnitudes at all epochs ($\pm 0.03$\,mag) was taken as the measurement uncertainty.

%___________________Table of RS Pup photometry
\begin{table}
\caption{Measured photometry of RS\,Pup.}
\label{photom_rspup}
\begin{tabular}{ccccc}
\hline
Date & MJD & phase & $m_B$ & $\sigma(m_B)$ \\
\hline
\noalign{\smallskip}
2006-10-28 & 54036.3592 & 0.6854 & 9.329 & 0.033 \\
2006-10-29 & 54037.3495 & 0.7093 & 9.343 & 0.038 \\
2006-12-20 & 54089.1463 & 0.9592 & 7.778 & 0.059 \\
2007-01-13 & 54113.2283 & 0.5404 & 8.970 & 0.040 \\
2007-01-22 & 54122.1317 & 0.7552 & 9.315 & 0.044 \\
2007-01-27 & 54127.1798 & 0.8770 & 8.564 & 0.037 \\
2007-03-24 & 54183.0114 & 0.2244 & 8.328 & 0.067 \\
\hline
\end{tabular}
\end{table}

\begin{figure}[ht]
\centering
\includegraphics[bb=0 0 360 288, width=8.5cm, angle=0]{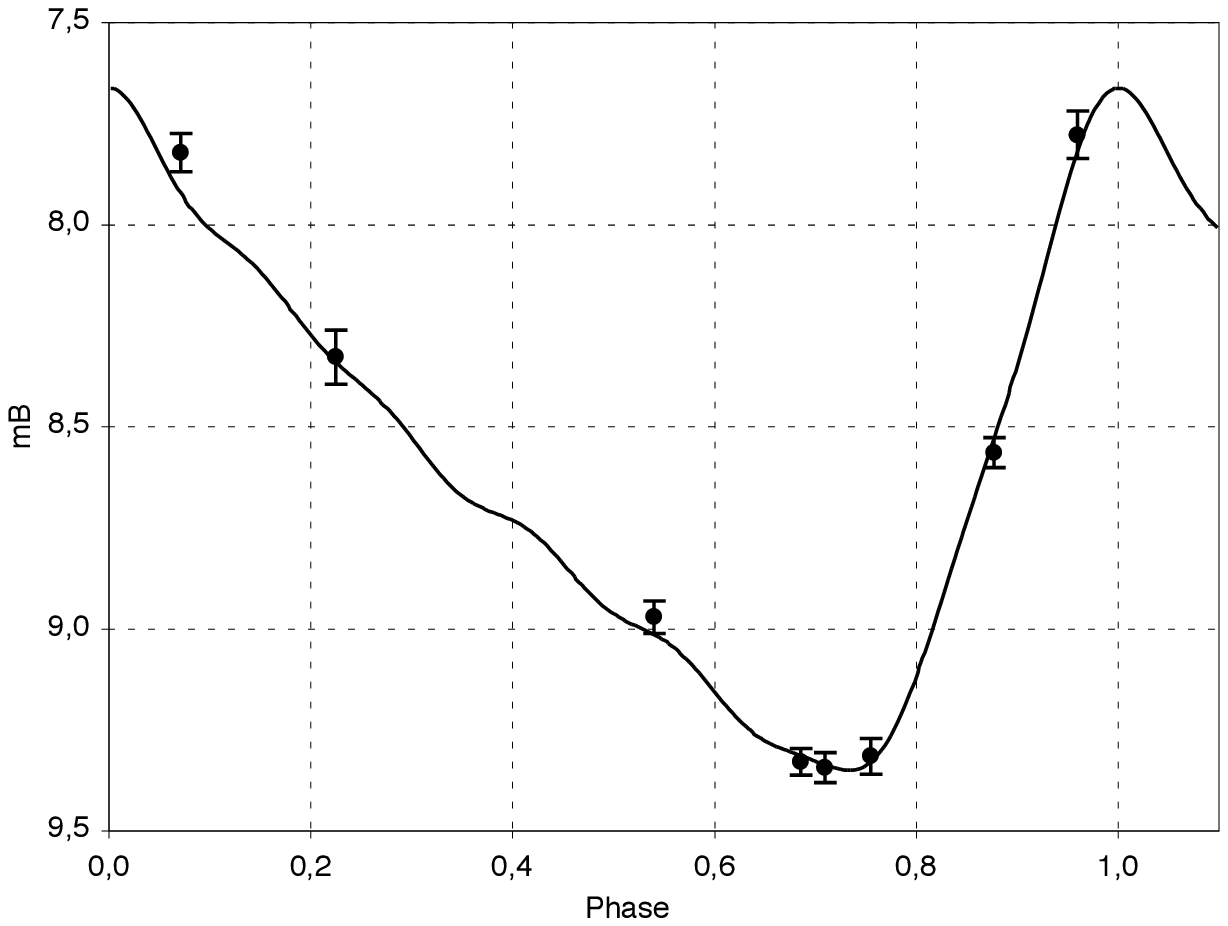}
\caption{Photometry of RS\,Pup from our observations, superimposed on its $B$ band light curve for $T_0 = \mathrm{MJD}\,54090.836$ and $P = 41.4389\,\mathrm{days}$.}
\label{RSPupPhotom}
\end{figure}

The PSF calibrator (HD\,70555) was chosen in the spectrophotometric catalogue assembled by Cohen et al.~(\cite{cohen99}). This gives us the possibility to use it as a convenient photometric reference to convert the measured ratio with RS\,Pup into magnitude. The {\it Hipparcos} catalogue (ESA~\cite{esa97}) gives $m_B = 6.249$. Although neither star was actually present on the detector, the measured photometry of RS\,Pup is relatively accurate (Table~\ref{photom_rspup} and Fig.~\ref{RSPupPhotom}).
From these magnitudes, we derive the following reference epoch of the maximum light:
$T_0 = \mathrm{MJD}\,54090.836 \pm 0.08$ (JD 2454091.336).
% = \mathrm{JD}\,2454091.336 \pm 0.08$$
For this fit, we adjusted a phase shift $\Delta \phi$ to match the measured photometric data points, using an interpolated light curve constructed from archival photometry (using Berdnikov's data\footnote{available from http://www.sai.msu.su/groups/cluster/CEP/PHE/}). We find no systematic magnitude shift down to a $\pm 0.02$\,mag level.

While the shape of the light curve has remained stable (which justifies the fit of the recent photometric data to the normal light curve), the period of RS\,Pup has varied dramatically over the last century. There are short term changes superimposed on a uniform secular period increase (to be discussed in a forthcoming paper). The instantaneous period of $P = 41.4389\,\mathrm{days}$ was derived from the most recent part of the ``Observed minus Calculated" ($O-C$) diagram, i.e., from the $O-C$ residuals calculated for the epochs between 2002 and 2006. We used this $(T_0, P)$ combination to phase our measurements.

%The period of RS\,Pup changed dramatically over the last century, and an extrapolation of its recent evolution gives an instantaneous period of $P = 41.4389\,\mathrm{days}$. We used this $(T_0, P)$ combination to phase our measurements.
%The behaviour of the pulsation period of RS\,Puppis and its implications will be discussed in a forthcoming paper.

%______________ Figure

After the PSF subtraction, the images of the night of the 20 December 2006 were deconvolved using the Lucy-Richardson algorithm in order to match the PSF FWHM with the other images in the series (normalized to a FWHM of 1.19$^{\prime\prime}$). After this procedure, we checked the consistency of the photometry of a sample of stars in the field and found a good agreement within less than $\pm 0.1$\,mag. Although the quality of the data of this particular night is low, this step is important in the context of nebular photometry to limit the contamination among neighbouring parts of the nebula.

\begin{figure*}[ht]
\centering
\includegraphics[width=8cm, angle=0]{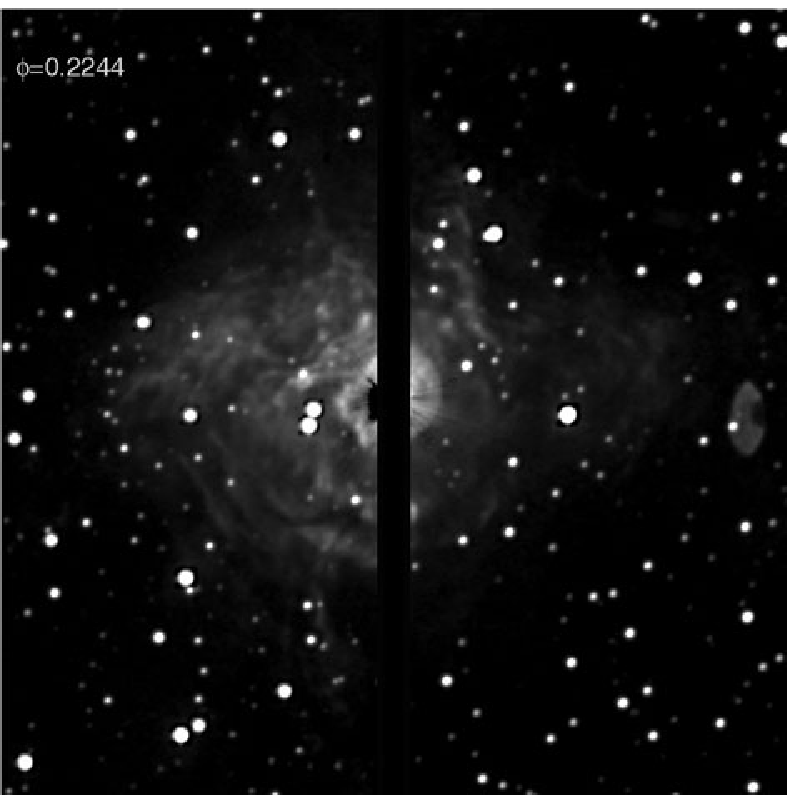}
\includegraphics[width=8cm, angle=0]{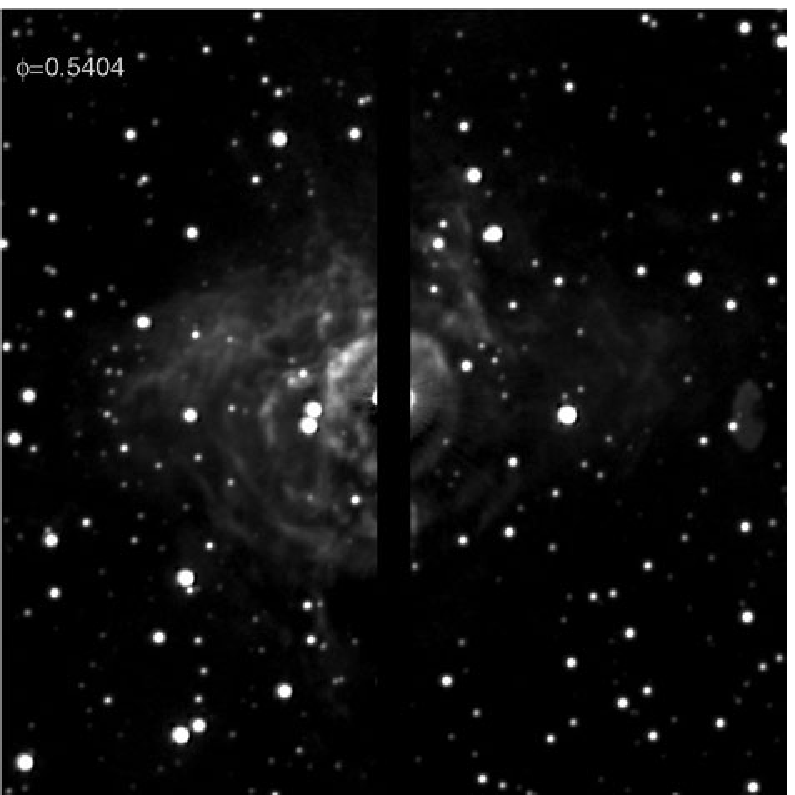}
\includegraphics[width=8cm, angle=0]{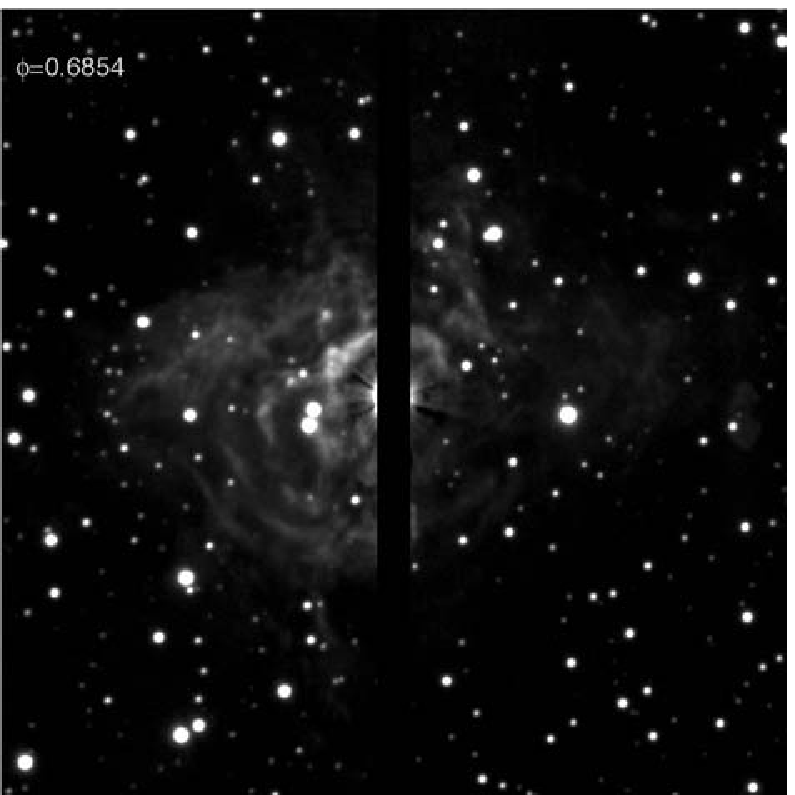}
\includegraphics[width=8cm, angle=0]{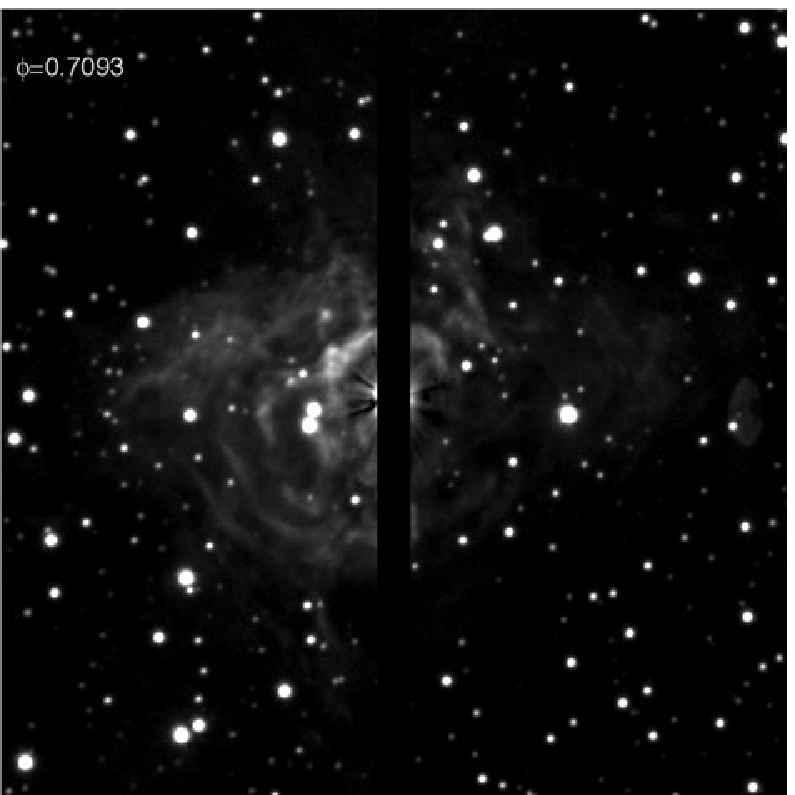}
\includegraphics[width=8cm, angle=0]{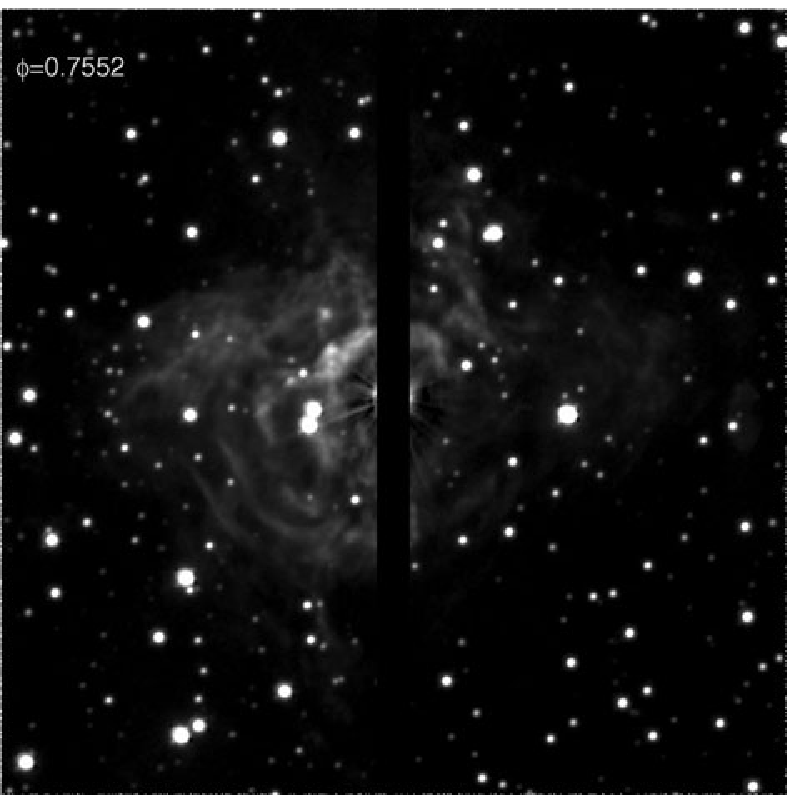}
\includegraphics[width=8cm, angle=0]{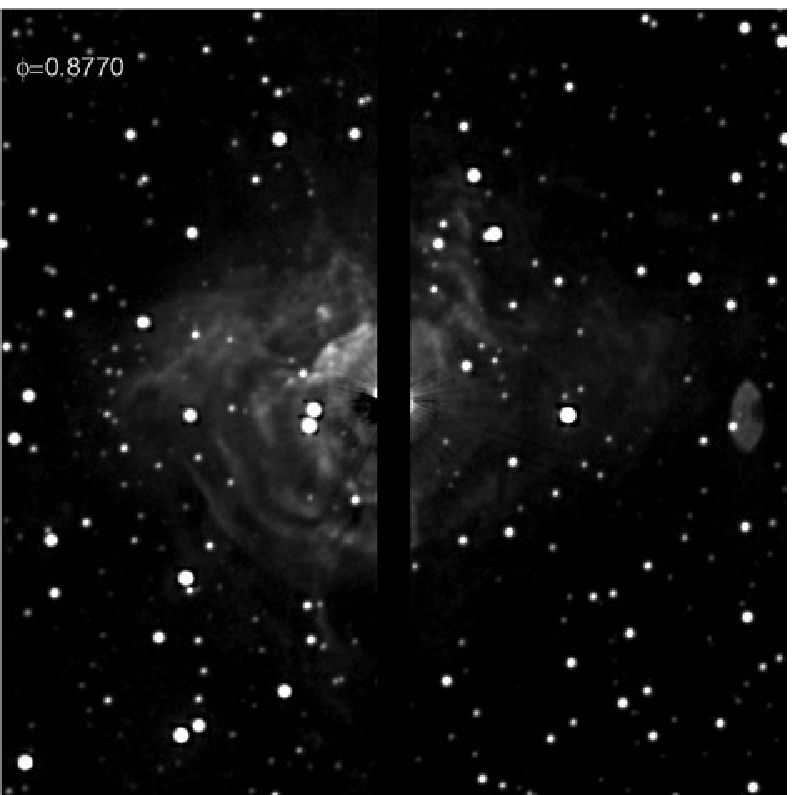}
\caption{PSF subtracted images of RS\,Pup, sequentially in phase from the upper left to the lower right. The intensity scale is identical for all images.}
\label{snapshots}
\end{figure*}

The variation of the aspect of the nebula over the six phases for which we have high quality images (excepting the night of 20 December 2006) is presented in Fig.~\ref{snapshots}. 

%______________ Figure

%__________________________________Distance
\section{Distance determination \label{distance}}

\subsection{Selection of the nebular features}

The determination of the distance to RS\,Pup following the method of Havlen~(\cite{havlen72a}) is based on the measurement of the phase difference between the variation of the star and the variation of isolated nebular features. When a particularly dense and localized feature can be isolated, we can make the reasonable hypothesis that the light diffused by this feature dominates the light diffused by the other superimposed layers. In other words, although several layers contribute to the local photometry (at the same projected position on the plane of the sky), the phase offset of the photometric variation can be considered identical to that of the localized ``blob".

%______________ Figure
\begin{figure}[ht]
\centering
\includegraphics[width=9cm, angle=0]{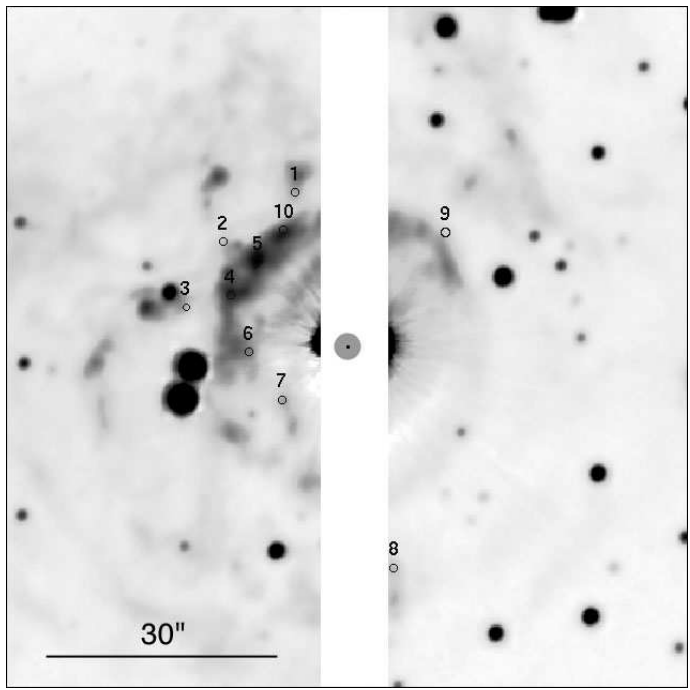}
\caption{Positions of the isolated nebular features chosen in the circumstellar nebula of RS\,Pup. Due to their phase-shifted variation cycles, not all features are simultaneously visible on this image, obtained on 13 January 2007 at a phase of $\phi = 0.5404$. The size of the circles marking each region corresponds approximately to the $5\times5$\,pixel boxes used to average the derived parameters.}
\label{blobs}
\end{figure}

The selection of the regions of interest was done in two steps:
\begin{enumerate}
\item identify visually all the well-defined compact nebular features that show a clear variation over several images,\\
\item select within this sample the features showing the largest photometric variation amplitude.
\end{enumerate}
Step 2 is necessary to select the features that are most likely to correspond to a single position in the nebula, thus avoiding the confusion of several superimposed layers located at different distances. This also selects the most accurate phase measurements. This procedure resulted in a selection of ten prominent nebular features. Figure~\ref{blobs} shows the position of these regions.
We chose features that are located as close as possible to the star, although avoiding the closest regions where residuals from the PSF subtraction are present. They are the most discriminating for the distance determination, as the integer number of cycles $N_i$ between their variation curve and the curve of the Cepheid is small. This translates into a large shift in distance for different values of $N_i$. 

\subsection{Photometry of the nebular features}

The measurement of the surface magnitude of an inhomogeneous nebula depends on the choice of aperture size. Choosing a too large aperture can lead to a degraded estimate of the variation of each individual component, while a too small aperture will produce inaccurate measurements due to the limited photometric signal-to-noise ratio (SNR). One advantage of the careful normalization of the PSF of our images before their photometric co-referencing is that our nebular photometry is mostly insensitive to the seeing fluctuations between each night.
We can therefore use each pixel of the nebula image individually to measure photometry, when the SNR is sufficient.
We thus did not adjust centroids on the nebular features, as they are highly variable, but instead we chose one fixed position for each of them over which we obtained the photometric measurements. The effective photometric aperture on each pixel therefore corresponds to the seeing PSF, with a constant FWHM of 1.19$^{\prime\prime}$ for all images (see Sect.~\ref{rspup_prep}).

In all the subsequent analysis of the nebula variation, the night of 20 December 2006 will be used only for verification purposes, and not for the fit itself. The deconvolution procedure that we used specifically for this night shows no indication of introducing a photometric bias on the photometry of the stellar sources, but could possibly create artefacts in the surface photometry of the nebula.

\subsection{Measured phase offsets \label{phase_offsets}}

The measured properties of the selected features and the amplitude and phase of their respective photometric variations are listed in Table~\ref{photom_blobs}.

\begin{figure*}[ht]
\centering
\includegraphics[width=6cm, angle=0]{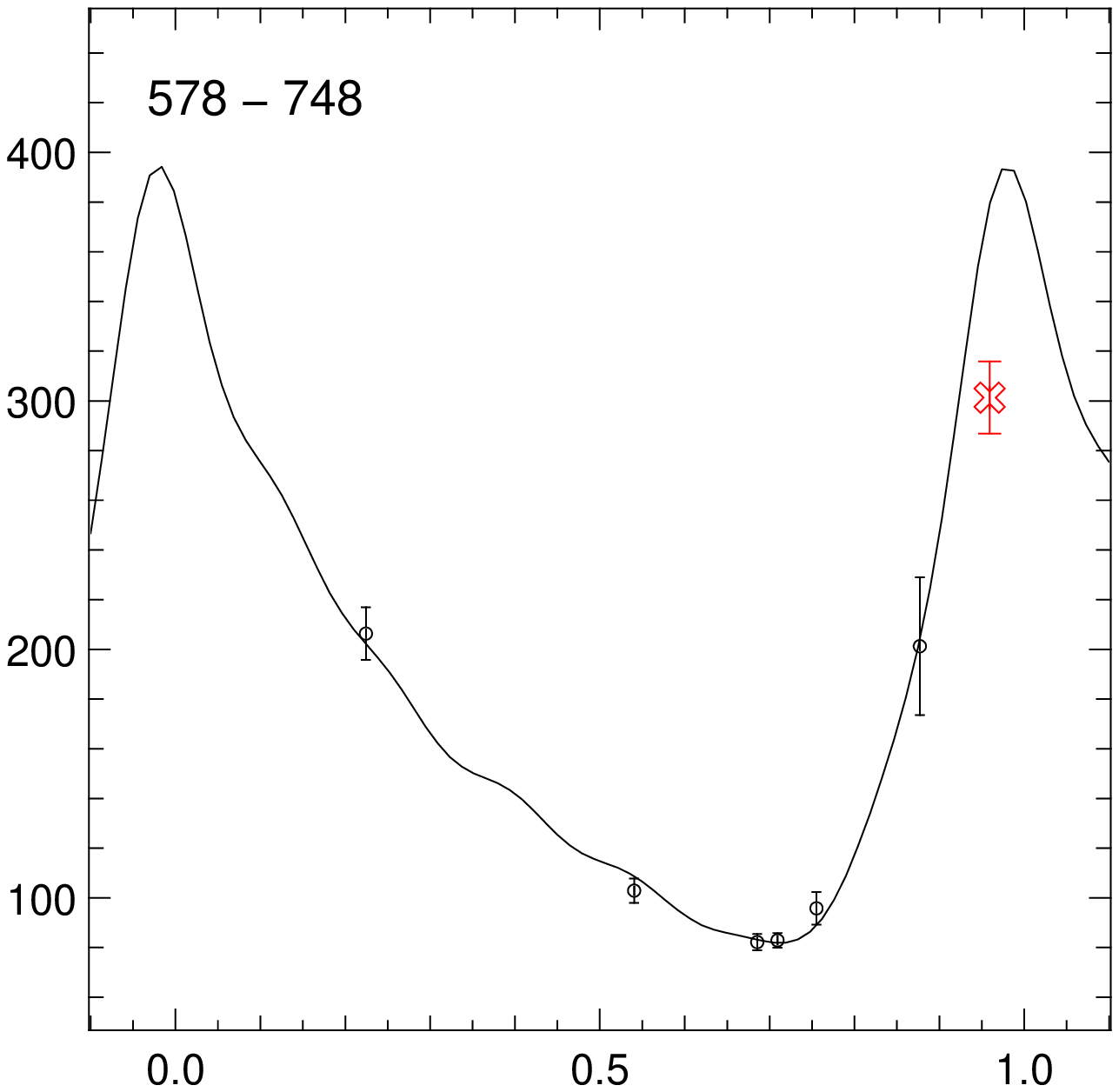}
\includegraphics[width=6cm, angle=0]{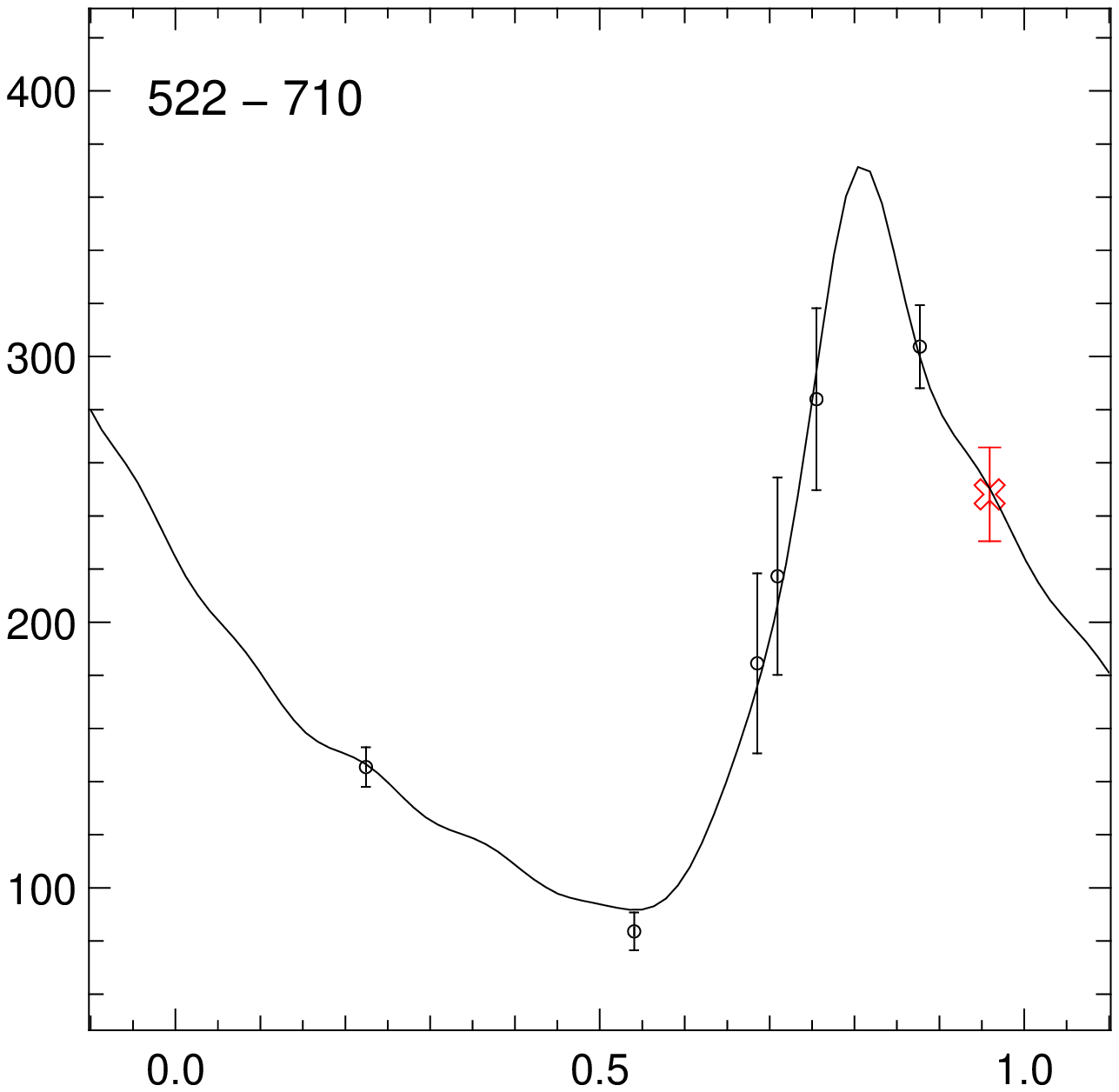}
\includegraphics[width=6cm, angle=0]{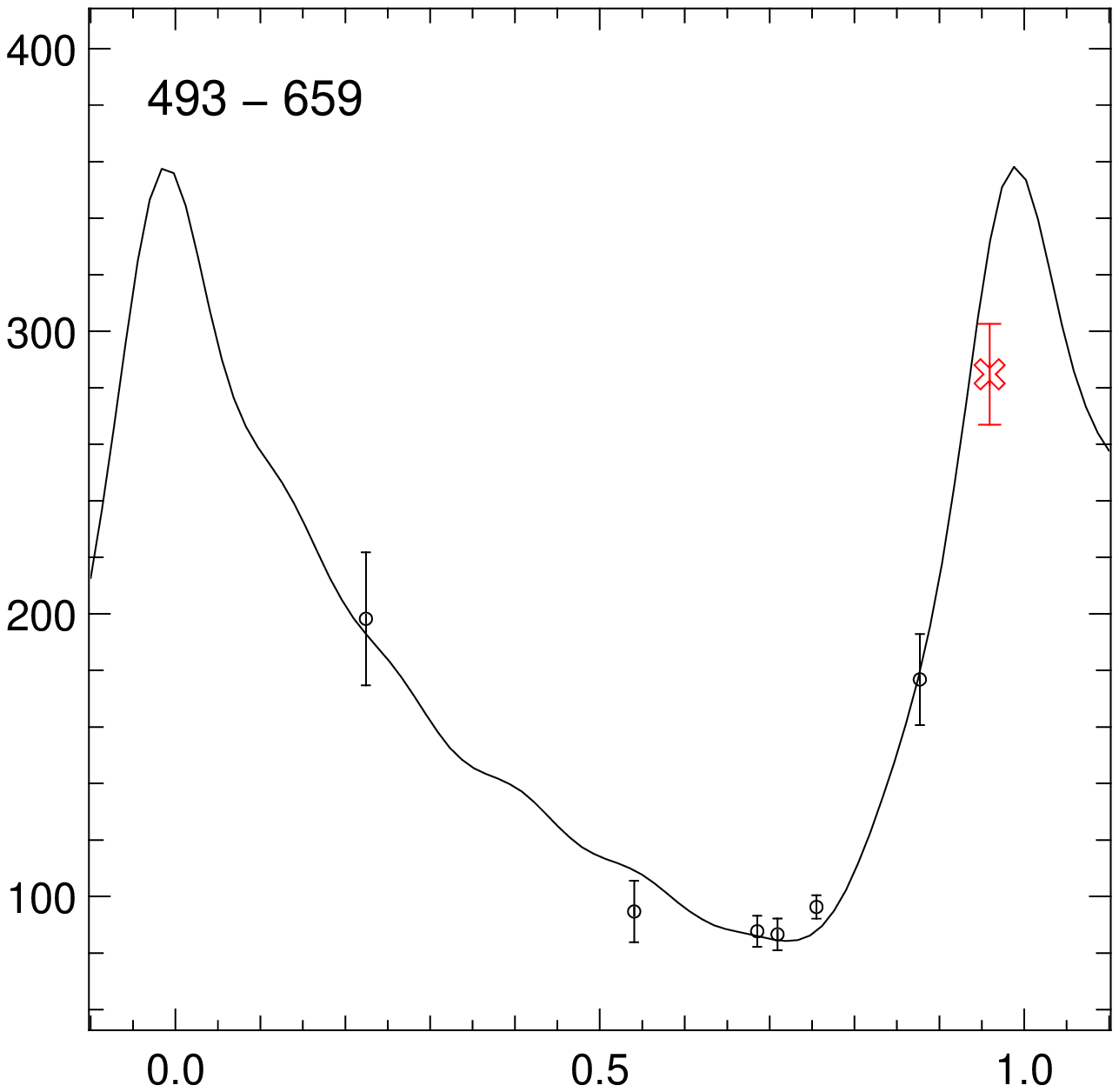}
\includegraphics[width=6cm, angle=0]{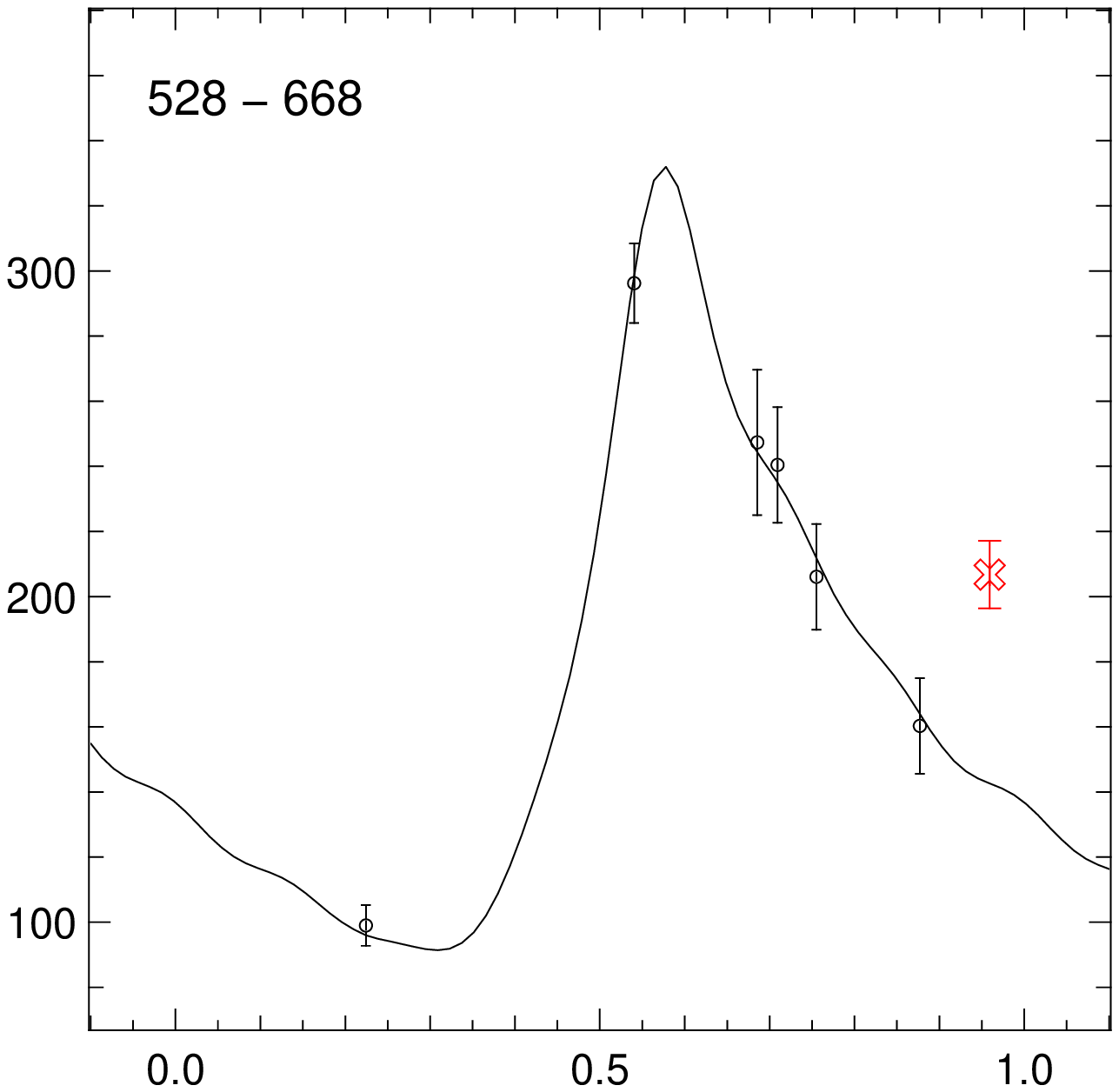}
\includegraphics[width=6cm, angle=0]{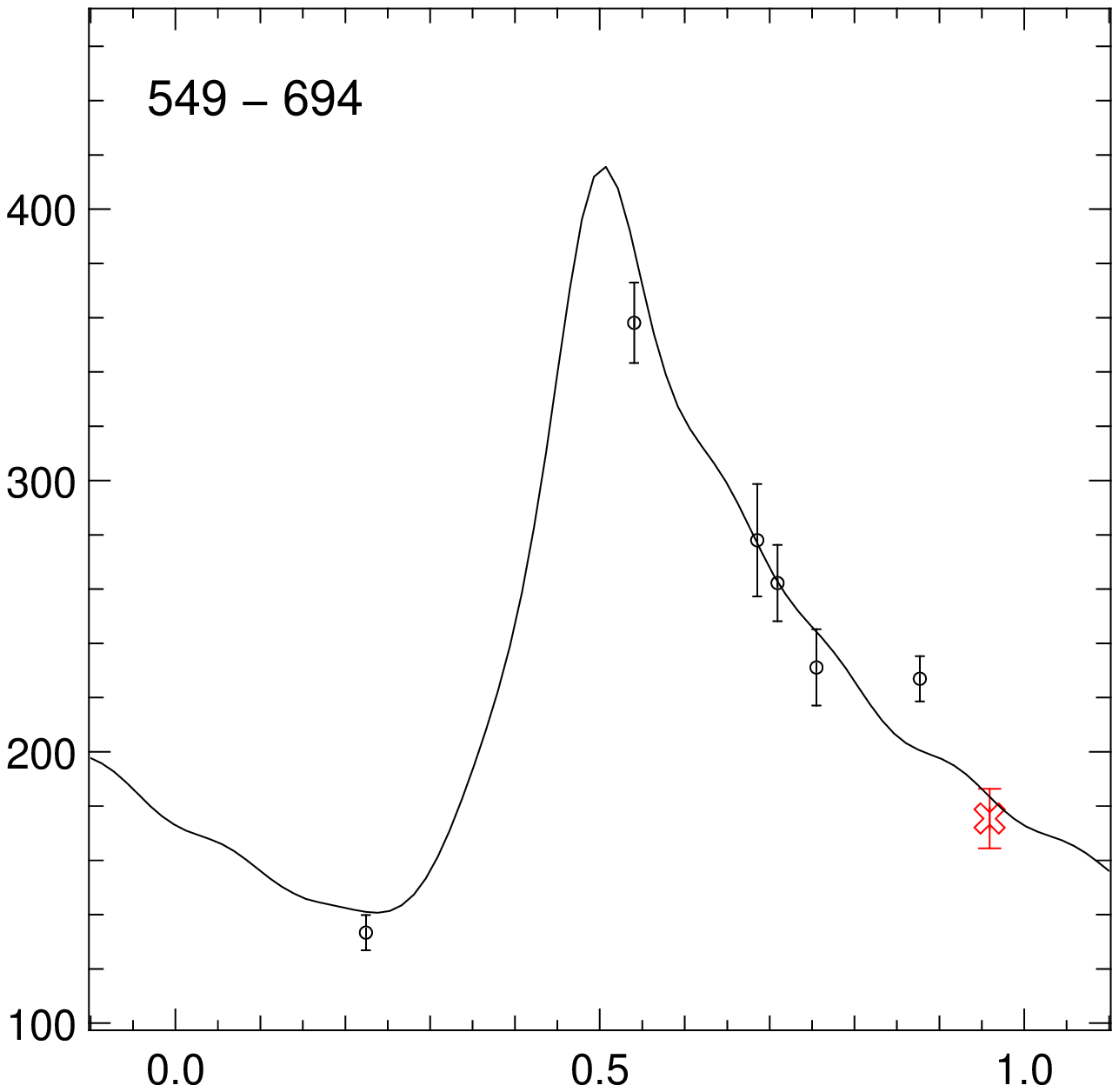}
\includegraphics[width=6cm, angle=0]{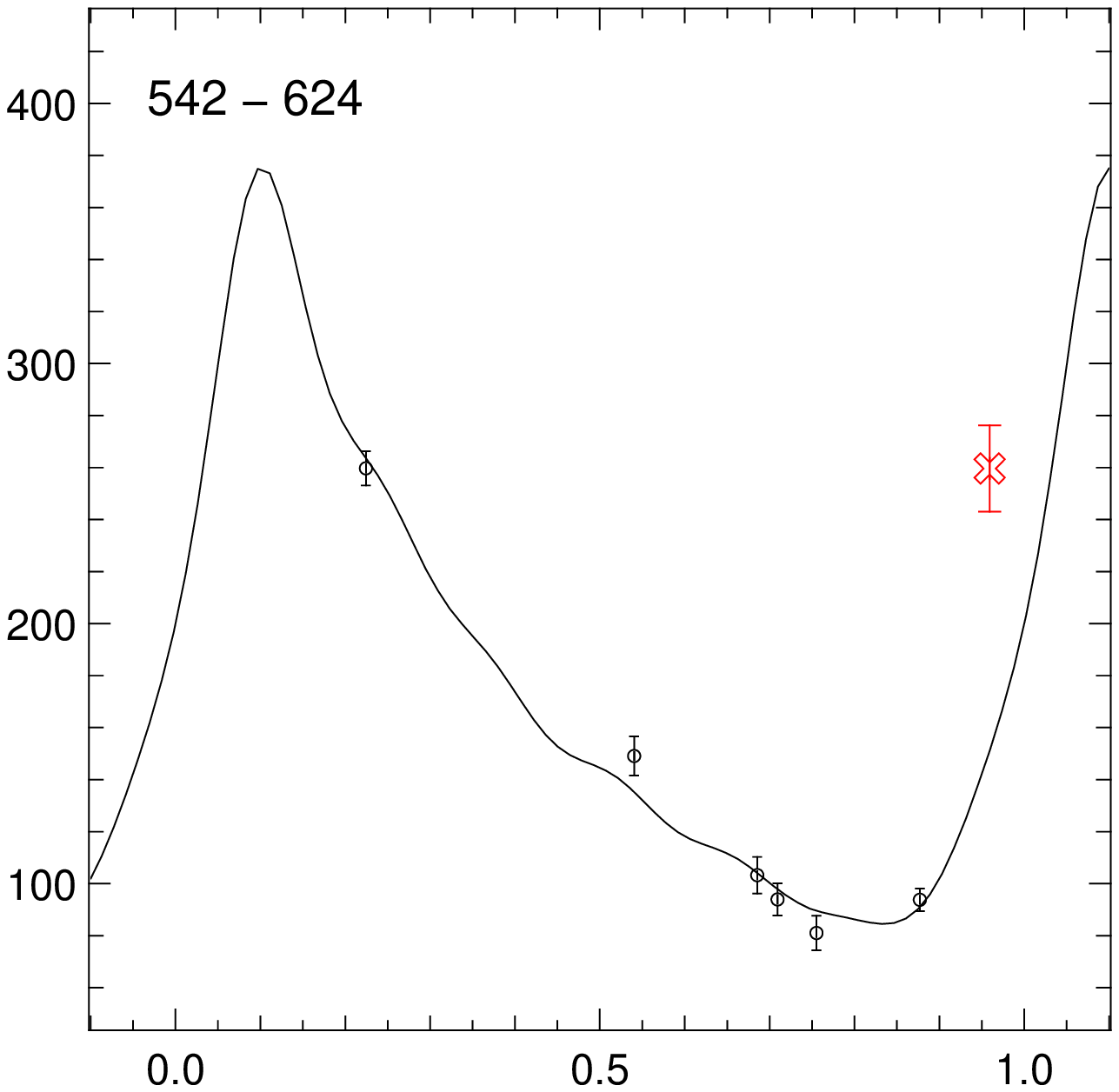}
\includegraphics[width=6cm, angle=0]{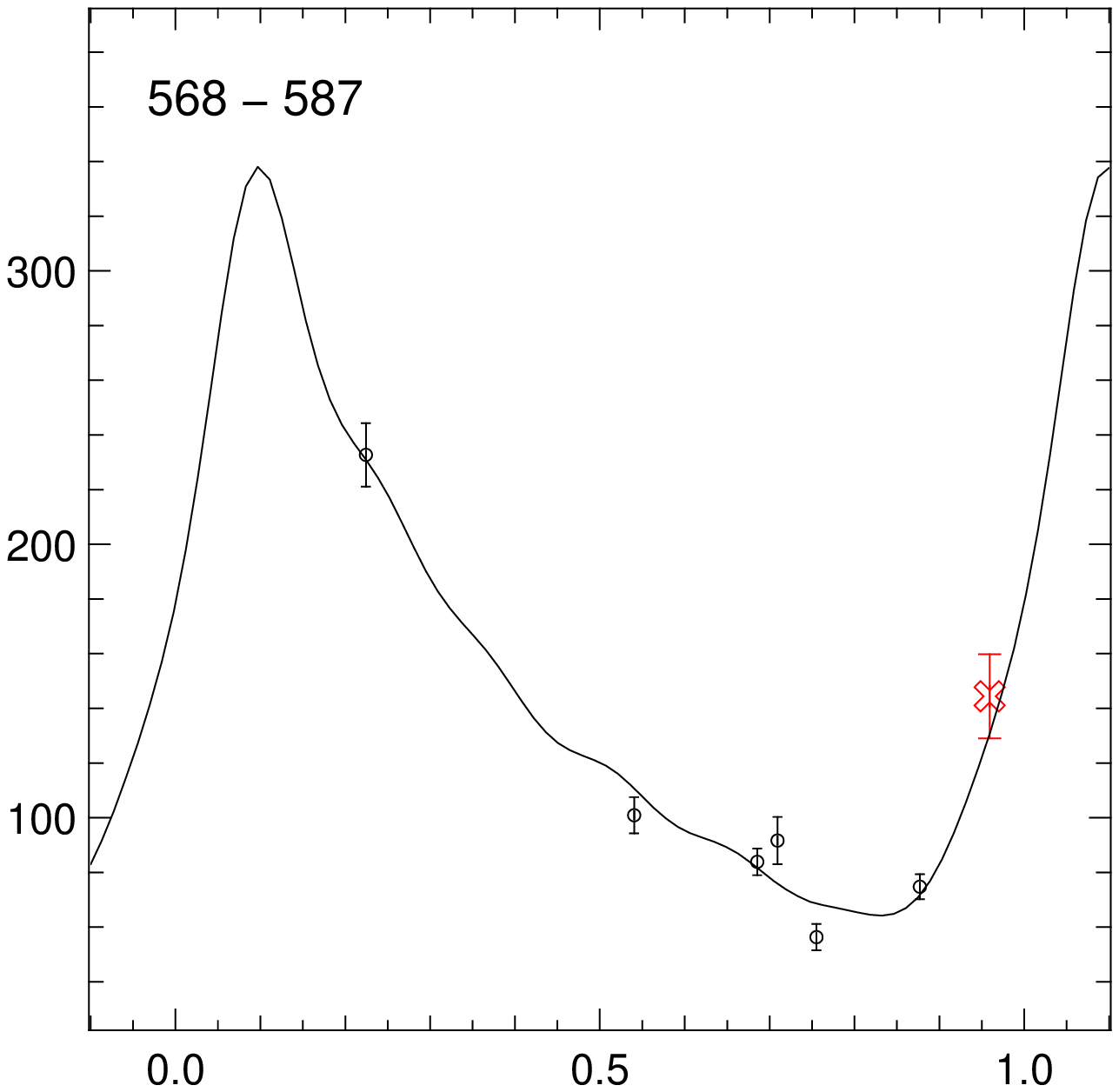}
\includegraphics[width=6cm, angle=0]{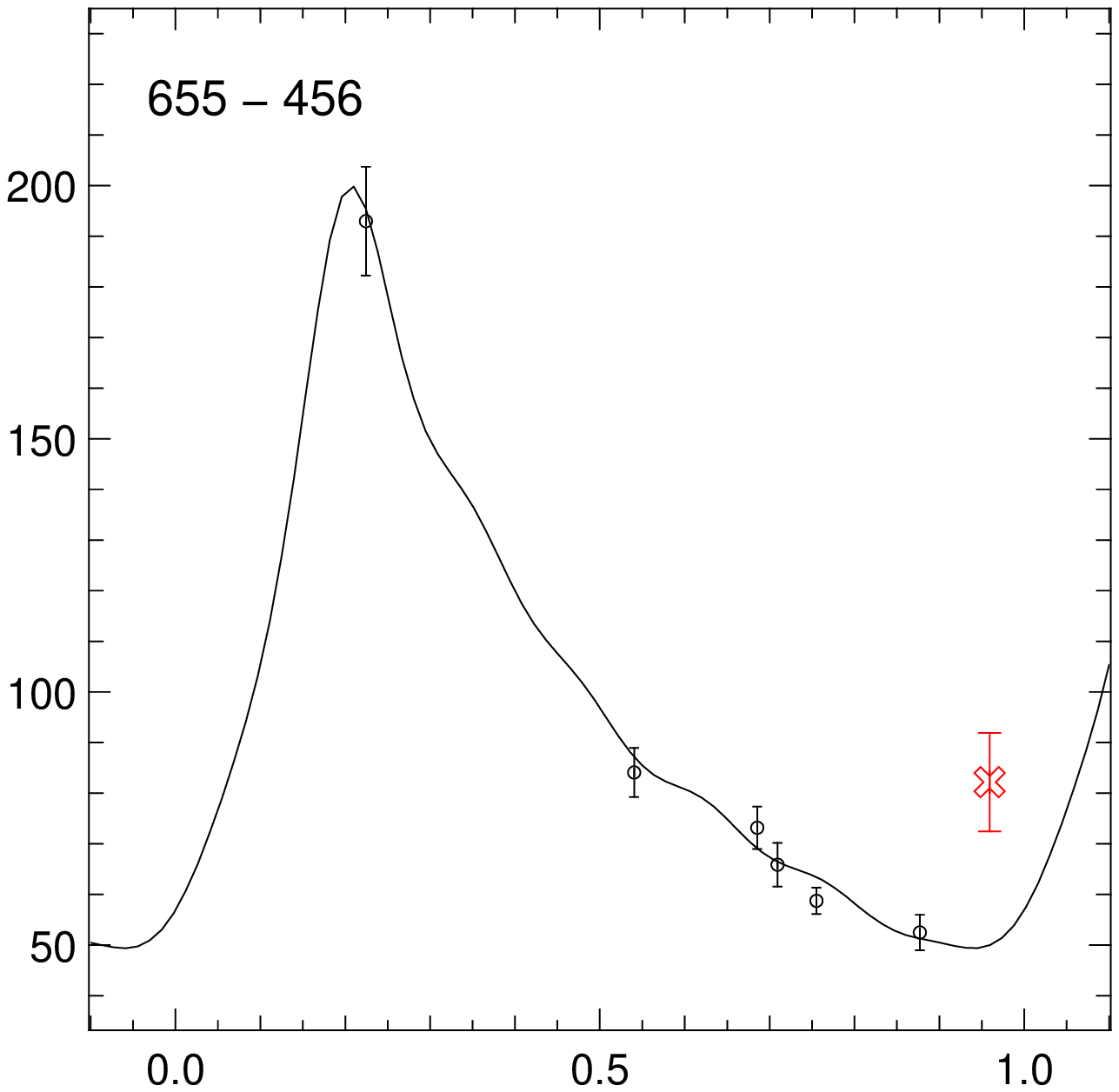}
\includegraphics[width=6cm, angle=0]{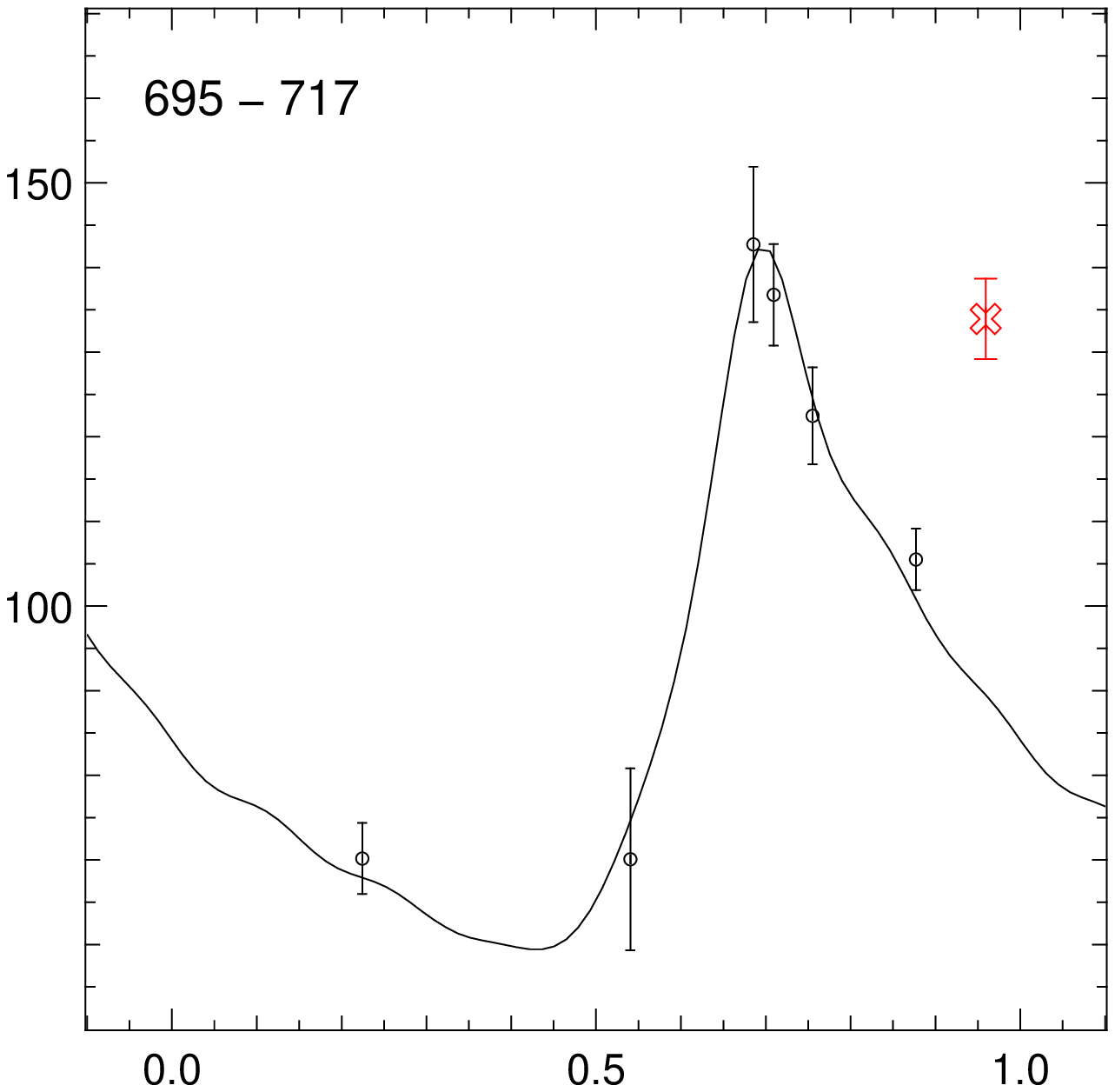}
\includegraphics[width=6cm, angle=0]{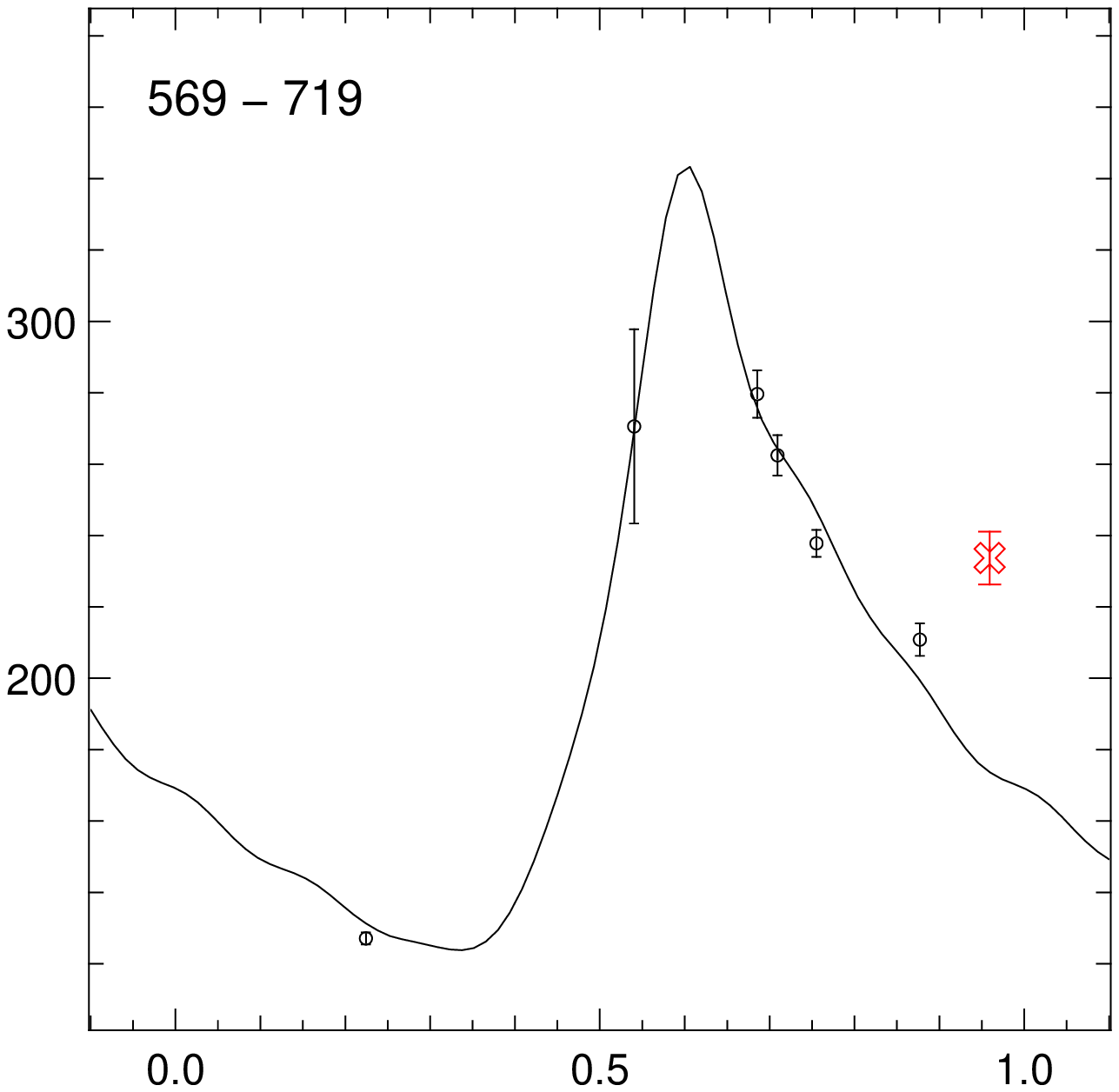}
\caption{Photometry of the ten nebular features used for the determination of the distance to RS\,Pup, with the corresponding adjusted photometric curves. The coordinates of each pixel in the images are indicated as ``$x-y$" in the upper left corner (See Table~\ref{photom_blobs} for further details). The horizontal axis is in phase units, and the vertical axis is in analog-digital units (ADUs).}
\label{phase-blobs}
\end{figure*}

The phases and amplitudes were subsequently averaged over a $5\times5$\,pixels box ($0.83^{\prime\prime}\times0.83^{\prime\prime}$) centered on each position. The fit was obtained by adjusting the phase, mean value and amplitude of the photometric curve of RS\,Pup to the photometry of each pixel using a classical $\chi^2$ minimization through a Levenberg-Marquardt algorithm. The fit was obtained separately on each pixel of the $5\times5$\,pixel boxes.
The standard deviation of the measurements over the $5\times5$\,pixels box, added quadratically to the fitting error bars was taken as its associated uncertainty. While the estimation of each phase offset is more precise, this conservative approach is intended to account for possible photometric contaminations by other nearby features.  The final listed phase and amplitude values correspond to the average of these parameters over these boxes. Figure~\ref{phase-blobs} shows the best fit curves for the selected nebular features detailed in Table~\ref{photom_blobs}. In this figure, the plotted error bars correspond to the standard deviation of the $5\times5$\,pixel window used to average the adjusted parameters. The data point marked with a cross was excluded from the fitting process as it corresponds to the night of 20 December 2006, when the telescope was accidentally out of focus.

It is interesting to remark that the feature \#1 selected by Havlen~(\cite{havlen72a}) corresponds approximately to our feature \#2, although an exact identification is made difficult by the absence of relative coordinates in his article (we used his Fig.~1). Havlen obtained a minimum light phase of $\phi = 0.52-0.57$, depending on the level of dilution of the photometric curve, while we obtain $\approx 0.55$ (Fig.~\ref{phase-blobs}). Considering the unknown uncertainty on Havlen's feature position, the agreement is satisfactory. The three other features he selected are located further away from RS\,Pup and do not correspond to ours.

%___________________Table of blobs photometry
\begin{table*}
\caption{Measured photometry of the selected nebular features. The $x$ and $y$ columns list the coordinates of the selected pixels in the data cube. The angular radius $\theta_i$ of these pixels is counted in arcseconds from RS\,Pup, and the azimuth is measured from north ($N=0^\circ$, $E=90^\circ$). The coordinates of RS\,Pup in the image, in pixels from the bottom left, are $x=619.3 \pm 0.2$, $y=628.2 \pm 0.2$. ``Mean$_i$" is the average value, and $A_i$ the amplitude of the adjusted photometric curve, expressed in ADU and $\Delta \phi_i$ is the phase offset with respect to RS\,Pup, with its associated total error (see text). $N_i$ is the computed integer number of periods between the variation cycle of RS\,Pup and that of the feature. $d_i$ is the derived distance to RS\,Pup. The six selected features for the final distance computation are marked with $*$ in the first column (see Sect.~Ê\ref{Ni_calc} for details).}
\label{photom_blobs}
\begin{tabular}{lccccccccl}
\hline
\# & $x$ (pix) & $y$ (pix) & $\theta_i$ ($^{\prime\prime}$) & Azimuth $(^\circ$) & Mean$_i$ (ADU) & $A_i$ (ADU)   & $\Delta \phi_i$ & $N_i$ & $d_i \pm \sigma$ (pc)  \\
\hline
\noalign{\smallskip}
1 & 578 & 748 & 21.10 & 139 & 185 $\pm$ 3 & 313 $\pm$ 11 & 0.983 $\pm$ 0.020 & 5 & 2034 $\pm$ 7 \\
2$*$ & 522 & 710 & 21.16 & 170 & 185 $\pm$ 15 & 280 $\pm$ 17 & 0.809 $\pm$ 0.012 & 5 & 1969 $\pm$ 5 \\
3$*$ & 493 & 659 & 21.65 & 196 & 175 $\pm$ 19 & 274 $\pm$ 43 & 0.989 $\pm$ 0.013 & 5 & 1985 $\pm$ 5 \\
4$*$ & 528 & 668 & 16.58 & 186 & 171 $\pm$ 11 & 241 $\pm$ 19 & 0.576 $\pm$ 0.009 & 4 & 1980 $\pm$ 4 \\
5$*$ & 549 & 694 & 16.03 & 167 & 232 $\pm$ 14 & 275 $\pm$ 28 & 0.504 $\pm$ 0.083 & 4 & 2016 $\pm$ 38 \\
6 & 542 & 624 & 12.89 & 213 & 181 $\pm$ 3 & 291 $\pm$ 20 & 0.102 $\pm$ 0.023 & 3 & 1726 $\pm$ 14 \\
7$*$ & 568 & 587 & 10.96 & 249 & 155 $\pm$ 9 & 274 $\pm$ 15 & 0.098 $\pm$ 0.023 & 3 & 2028 $\pm$ 16 \\
8$*$ & 655 & 456 & 29.28 & 312 & 99 $\pm$ 9 & 151 $\pm$ 20 & 0.207 $\pm$ 0.036 & 8 & 2010 $\pm$ 9 \\
9 & 695 & 717 & 19.42 & 80 & 87 $\pm$ 6 & 83 $\pm$ 4 & 0.697 $\pm$ 0.016 & 5 & 2104 $\pm$ 6 \\
10 & 569 & 719 & 17.28 & 149 & 197 $\pm$ 6 & 220 $\pm$ 9 & 0.602 $\pm$ 0.014 & 4 & 1910 $\pm$ 6 \\
\hline
\end{tabular}
\end{table*}

%______________ Figure

\subsection{Distance estimation and statistical uncertainty \label{Ni_calc}}

\begin{figure}[ht]
\centering
\includegraphics[bb=0 0 360 288, width=9cm, angle=0]{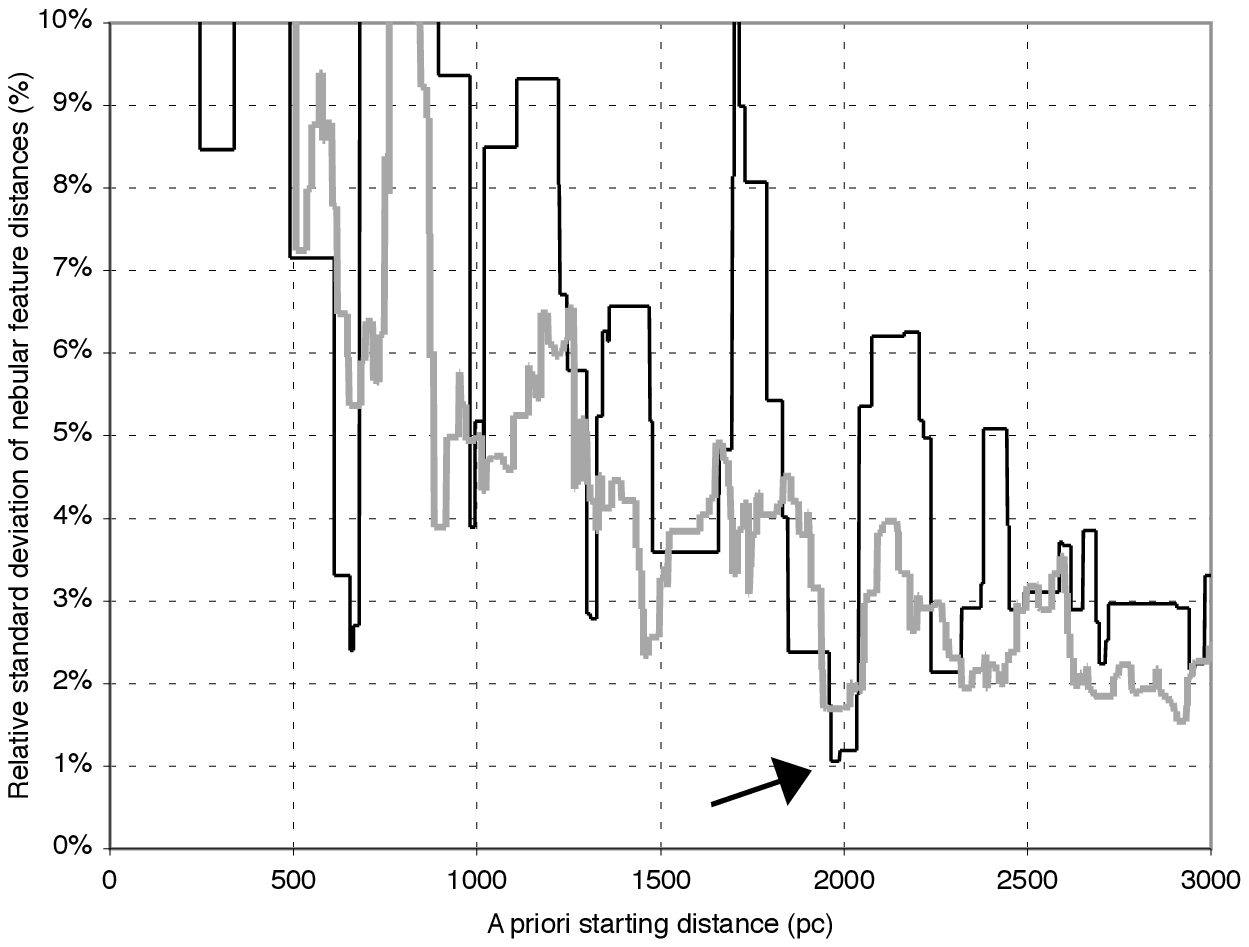}
\caption{Relative dispersion of the $d_i$ distance estimates as a function of the initial assumed distance to RS\,Pup for 10 (thin black curve) or 21 (thick grey curve) nebular features. The arrow points at the minimum dispersion of 1.1\%.}
\label{disp-di-blobs}
\end{figure}

We assume in the following that the integer number of periods $N_i$ separating the photometric variation of RS\,Pup from the variation of each nebular feature is increasing with the angular distance from the star. This is a natural consequence of our assumption that the selected features are located close to the plane of the sky (see discussion above). This means that a nebular feature with index $i$ located closer to RS\,Pup than another region $j$ will have $N_i \le N_j$.
From the phase offset $\Delta \phi_i$ of each feature (with $1 \le i \le 8$),  we can derive a distance $d_i$ to RS\,Pup assuming an arbitrary value for $N_i$, from the simple relation:
$$ d_i = \frac{(N_i + \Delta \phi_i)\,P}{5.7755\,10^{-3}\ \theta_i}$$
where $d_i$ is in parsecs, $P$ in days, and $\theta_i$ in arcseconds. As rightly pointed out by Havlen~(\cite{havlen72a}), it is not possible to obtain a unique distance solution, independently of the number of measured features. However, for the right values of the $N_i$ numbers, we expect a marked convergence of the estimated $d_i$ towards a single value. As we assumed that $N_i$ is increasing with the separation from RS\,Pup, we can summarize the determination of the distance to a study of the relative standard deviation of the $d_i$ distances as a function of the {\it a priori} assumed $d$ distance to RS\,Pup. Figure~\ref{disp-di-blobs} clearly shows the best convergence for a starting value around 1\,990\,pc. The best-fit unweighted average distance $d$ of all $d_i$ in this case is $d = 1\,992 \pm 20\,\mathrm{pc.}$ If we apply a weighting of the $d_i$ measurements by the inverse of the phase variance of the corresponding nebular feature, we obtain 1\,983\,pc, within half a $\sigma$ of the previous value. The unweighted median distance is 1\,997\,pc.

The individual formal error bar on each $d_i$ value is usually below 1\%, and in some cases even better than 0.3\%, but their dispersion is larger due to their non-planar spatial distribution around RS\,Pup. To take this into account, the uncertainty on the average distance was computed using the unweighted bootstrapped standard deviation of all $d_i$, minus the two highest and two lowest values of our set.
A description of the bootstrapping technique can be found in Appendix~B of Kervella et al.~(\cite{kervella04}). This clipping of four $d_i$ distance values (out of ten) is useful to reduce our sensitivity to the dispersion of the measured blobs relatively to the plane of the sky. Features located closer or further from us than RS\,Pup will have a smaller apparent angular separation from RS\,Pup, and the distance estimate will be uncertain. We suspect this is the case for the nebular feature \#6 in Table~\ref{photom_blobs}. However, fits with all ten blobs included, or inversely with six features clipped out, leads to exactly the same unweighted average value of $1\,992 \pm 19\,\mathrm{pc}$, so this clipping does not introduce any significant bias.

%______________ Figure
\begin{figure}[ht]
\centering
\includegraphics[bb=0 0 360 288, width=9cm, angle=0]{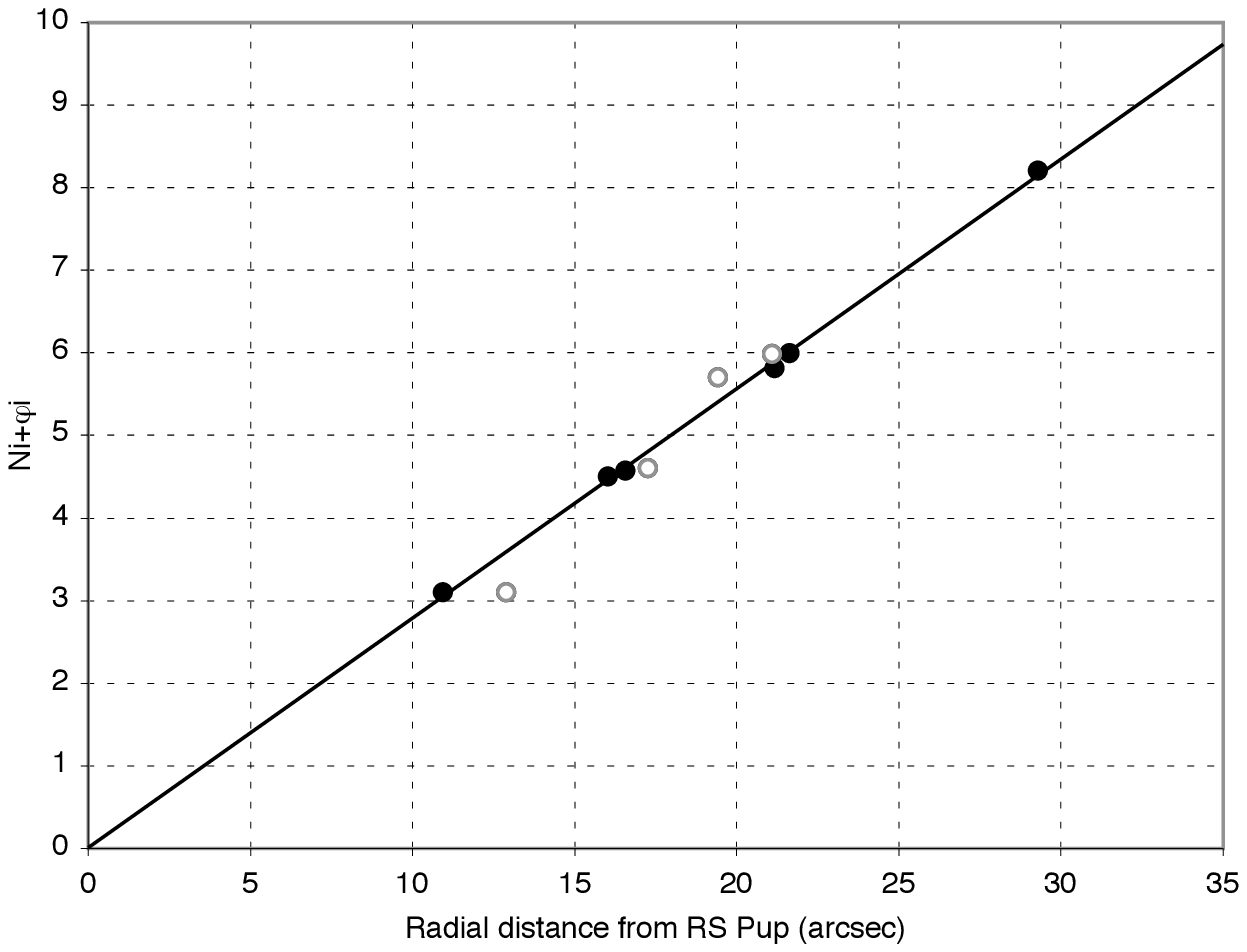}
\caption{Total phase $N_i + \phi_i$ of the observed variations of the nebular features around RS\,Pup as a function of their radial angular distance from the star for the $N_i$ values given in Table~\ref{photom_blobs}. The line is a two-parameter linear fit through the points (see text). The data points shown as open circles were excluded from the fit (two largest and two smallest distance estimates from the set).}
\label{dist-fit}
\end{figure}

As shown in Fig.~\ref{disp-di-blobs}, other starting distances than 1\,990\,pc provide some convergence, in particular around 650\,pc and 1\,340\,pc, but the dispersion of the $d_i$ is much larger than for 1\,992\,pc (2.4\% and 2.8\%, compared to 1.1\%). They correspond to different sets of increasing $N_i$ values. The corresponding reduced $\chi^2$ are respectively $\chi^2_{\rm red} = 73$, 775 and 151, for the three increasing distances. The smaller $\chi^2_{\rm red}$ for 650\,pc is due to two effects: 
\begin{itemize}
\item The error bars on each phase offset $\phi_i$ translate into constant uncertainties on the $d_i$ values, independently of the chosen $N_i$. If the average distance becomes smaller (by reducing all the $N_i$ simultaneously for instance), the \emph{relative} error of each $d_i$ increases, while the errors themselves remain constant. As a consequence, for the smaller average distance, the $\chi^2$ becomes smaller, while the standard deviation of the $d_i$ remains the same. The $\chi^2$ is therefore a poor indicator of the convergence of the method, compared to the straight standard deviation.
\item The 650\,pc distance is nearly a third of 1\,992\,pc, and it betrays a geometrical aliasing effect of the distribution of the selected blobs on the nebula. We can exclude this small distance (as well as 1\,340\,pc) using our verification fit with a distinct set of 21 nebular features (see also Sect.~\ref{verif_fit}), which gives $\chi^2_{\rm red}=924$ around 650-700\,pc compared to $\chi^2_{\rm red}=674$ around 2\,000\,pc.
\end{itemize}

One should also note that a distance of exactly twice 1\,992\,pc is also a possibility. This is a natural consequence of the method, as the number $N_i$ of cycles is exactly doubled for each blob, but such a value of $\approx 4\,000$\,pc is excluded by previous determinations from other methods (see Sect.~\ref{discussion}). This also applies to larger integer multiples.

We can check {\it a posteriori} that the derived distance is self-consistent by fitting the derived total phases $N_i + \phi_i$ as a function of the angular radius from RS\,Pup for both the slope and the zero point. The slope is proportional to the inverse of the distance, and the zero point should give a null phase. The derived distance is 1\,993\,pc, and the residual phase is very small ($\Delta \phi_0 = 0.007$) for the selected features (Fig.~\ref{dist-fit}). The agreement is not so good for the other distances where a local convergence is observed, giving further confidence that they are not correct.

\subsection{Systematic error sources and final distance}

\subsubsection{Astrometry and photometry}

The pixel scale of the EMMI detector is $0.1665 \pm 0.0006^{\prime\prime}$.pix$^{-1}$, resulting in a systematic uncertainty of 0.4\% on all $\theta_i$ values, and consequently on the distance. The position of RS\,Pup was determined with an accuracy of $\pm 0.2$\,pixel on the reference image, from the intersection of the large diffraction spikes created by the secondary mirror support (visible in Fig.~\ref{HD70555psf}, bottom). This translates into a relative uncertainty of 0,033$^{\prime\prime}$ on each measured $\theta_i$. For the smallest angular distance of feature \#7 ($\theta_i = 10.96^{\prime\prime}$), this corresponds to a relative uncertainty of 0.3\%.

The angular separations $\theta_i$ themselves do not contribute to a systematic uncertainty as we considered pixel to pixel photometry, without centroiding the features. The accuracy of the registration of the images with respect to each other is very high as we used a cross-correlation over hundreds of background stars (star positions are constant on the detector). The photometric normalization of each image with respect to the first in the series was achieved using a sample of field stars with an accuracy of a few 0.01\,mag (Sect.~\ref{rspup_prep}). The subtraction of the telescope PSF introduces additional photometric errors, but they are taken into account in this statistical uncertainty. More generally, the photometry is not expected to be a significant contributor to the overall systematic uncertainty, as the residual fluctuations are reflected into the statistical phase uncertainties listed in Table~\ref{photom_blobs}.

\subsubsection{Period and reference epoch}

The computation of the phase offset relies on the assumed $(P,T_0)$ pair. An error on these two values will have an impact on the determined phase offset of a given feature, and thus on its estimated linear distance from RS\,Pup. We derived the time of the actual maximum light epoch $T_0$ from our own data, and thus obtained a value suitable for the epoch of our observations. An uncertainty of 0.08\,day is attached to this estimate. It translates into a maximum systematic uncertainty of $0.08/41.4 = 0.2\%$ on the distance. As they are spread over 147\,days, our observations cover about 3.5\,periods of RS\,Pup. Assuming an uncertainty of $\pm 0.02$\,day on the period $P$, this will create a maximum $\pm 0.07$\,day uncertainty on the last maximum dates, equivalent to $\pm 0.2$\% on the distance.

\subsubsection{Integer number of periods $N_i$ and scattering}

The computation of the distance for each nebular region is based on the hypothesis that they are located in the plane of the sky, at the same distance as RS\,Pup itself. This is underpinned by the reasonable assumption that the distribution of the features around the star is, on an average, isotropic. Several features could be located closer or farther from us than RS\,Pup, resulting in a larger or smaller derived distance from these particular regions, depending on the integer rounding error made on the $N_i$ values. However, it is unlikely that it is simultaneously the case for all features, especially as we consider features located close to the star, hence with low $N_i$ values (Table~\ref{photom_blobs}).
%In addition, Havlen~(\cite{havlen72a}) indicates that the scattering process on dust grains is limited to a range of approximately $\pm 15^\circ$ from the plane of the sky, thus leading to an uncertainty of only 3-4\% on the computed distance corresponding to each region.
Considering that this uncertainty due to $N_i$ is of statistical nature (averaging out with several blobs), we do not introduce here a specific systematic error contribution. The statistical dispersion of the measurements is included in the bootstrapped error bar given in Sect.~\ref{Ni_calc}.

%\subsubsection{Scattering angle}

The scattering angle of the light from the Cepheid by the dust can in principle create an uncertainty on light echo distances. This question is discussed extensively by Havlen~(\cite{havlen72b}). Although this is true in the case of scattering by a continuous dusty medium, we rely in the present work on spatially isolated, well defined nebular features. This makes our distance estimate mostly insensitive to the scattering law. At the limit, the scattering by an unresolved, isolated compact dust region would be fully independent of the scattering function. For this reason, we do not introduce a specific systematic uncertainty due to the unknown scattering function.

\subsubsection{Choice of the nebular features \label{verif_fit}}

The choice of the nebular features and their number could possibly create a bias. As a verification, we considered a distinct sample of 21 positions in the nebula, extending up to $\theta_i = 40^{\prime\prime}$ from RS\,Pup. We derived from this sample an unweighted average distance of $2\,025 \pm 18$\,pc (Fig.~\ref{disp-di-blobs}, thick grey curve), only 1$\sigma$ away from the distance obtained in Sect.~\ref{Ni_calc} ($1\,992 \pm 20$\,pc).
%Due to the distribution of values, the error bar does not decrease although the number of blobs is larger than our reference sample (bootstrapped error bars).

The slightly different distance value is due to the inclusion of poorly defined nebular features that show a ``smoothed" photometric variation curve due to the superimposition of several layers located on the line of sight. The estimation of the phase for these features is thus slightly biased, as shown by Havlen~(\cite{havlen72a}). An indication that this bias is present is that the unweighted median distance from this large sample is $2\,012 \pm 22$\,pc, closer to the value obtained with the selected features. Moreover, the zero point of the retrospective two-parameter fit of the blob distances gives $\Delta \phi = -0.10$, significantly worse than for our carefully selected reference sample.

From this result, we affect a systematic uncertainty of $\pm 0.8$\% ($\pm 16$\,pc) to the choice of the nebular features.

\subsubsection{Final distance and uncertainty}

As discussed in the previous section, several systematic uncertainty sources have an impact on the distance of RS\,Pup determined from its light echoes. Table~\ref{systematics} lists the most important ones, with their estimated relative impact on the determined distance. All these terms have a linear effect on the measured distance. 
Adding quadratically all these contributions, we obtain a total systematic uncertainty of 1.0\% (Table~\ref{systematics}). This small number shows that the light echo method is very robust in the case of a periodic star with a well-defined variation curve. In particular, the selection of the nebular features, although an important point, is not a strong limitation to the final accuracy. Writing separately the statistical and systematic uncertainties, the final distance to RS\,Pup is therefore:
%-------
$$d = 1\,992 \pm 20 \pm 20\,\mathrm{pc.}$$
%-------
corresponding to a parallax of $\pi = 0.502 \pm 0.007$\,mas (total relative uncertainty of 1.4\%) and a distance modulus of $\mu = 11.50 \pm 0.03$\,mag.

%___________________Table of systematic uncertainties
\begin{table}
\caption{Systematic uncertainty contributors on the distance of RS\,Pup.}
\label{systematics}
\begin{tabular}{lcc}
\hline
Term & Value & $\sigma_{\rm syst}$ \\
\hline
\noalign{\smallskip}
Pixel scale & $0.1665 \pm 0.0006^{\prime\prime}$ & 0.4\% \\
\noalign{\smallskip}
Position of RS\,Pup & $\pm 0.2$\,pixel & 0.3\% \\
\noalign{\smallskip}
Reference epoch & $\mathrm{MJD}\,54090.836 \pm 0.08$ & 0.2\% \\
\noalign{\smallskip}
Photometric period & $41.4389 \pm 0.02\,\mathrm{days}$ & 0.2\% \\
\noalign{\smallskip}
Choice of features & $\pm 16$\,pc (see text) & 0.8\% \\
\noalign{\smallskip}
\hline
\noalign{\smallskip}
Total & $\pm 20$\,pc & 1.0\% \\
\noalign{\smallskip}
\hline
\end{tabular}
\end{table}

%__________________________________Discussion
\section{Comparison with existing distance estimates \label{discussion}}

\subsection{Trigonometric parallax}

Due to its large distance, the trigonometric parallax of RS\,Pup measured by Hipparcos cannot be accurate. The original Hipparcos catalogue published a value of $\pi = 0.49 \pm 0.68$\,mas (ESA~\cite{esa97}), which is not significant. However, a revised parallax was published by van Leeuwen et al. (\cite{vanleeuwen07c}) as $\pi = 1.44 \pm 0.51$\,mas, corresponding to a 3$\sigma$ detection, and the final value that appears in the revised Hipparcos parallax catalogue is $\pi = 1.92 \pm 0.65$\,mas (van Leeuwen~\cite{vanleeuwen07a}; see also van Leeuwen~\cite{vanleeuwen07b}). Given the limited accuracy of Hipparcos ($\sim 1$~mas), we do not have an explanation for this apparent detection of parallax, in disagreement with our determination.

\subsection{Infrared surface brightness}

A more promising distance determination comes from the infrared surface brightness method, a variant of the Baade-Wesselink technique. Barnes et al. (\cite{barnes05}) included RS\,Pup in their catalogue, with a Bayesian distance of $1\,984 \pm 122$ pc and a least squares distance of $2\,004 \pm 59$\,pc, both in excellent agreement with the present determination, although less accurate. However, a systematic uncertainty arises with this technique (not accounted for in the above error bars), due to the projection factor used to convert radial velocities to pulsation velocities. Barnes et al. used a classical $p$-factor first derived by Gieren et al.~(\cite{gieren93}), namely $p = 1.39 - 0.03 \log P$, which amounts to 1.341 in the case of RS\,Pup ($\log P = 1.6174$). Fouqu\'e et al.~(\cite{fouque07}) derived a distance of 1\,830\,pc based on the same data, but using a lower projection factor based on a mean relation derived by Nardetto et al.~(\cite{nardetto07}), $p = 1.366 - 0.075 \log P$ (cross-correlation method), which leads to 1.245 in the case of RS\,Pup. This revision is based on recent hydrodynamical models of RS\,Pup itself, combined with high-resolution spectroscopic measurements. The question of the $p$-factor will be further discussed in a forthcoming paper. The interested reader can refer to Groenewegen's~(\cite{groenewegen07}) recent work.

In conclusion, we can regard the infrared surface brightness distance as supporting our result, although both its random and systematic uncertainties are larger than those of the present geometric distance measurement. For the sake of better visualization, Table~\ref{distanceobs_rspup} gives a summary of the distance values of RS\,Pup obtained from various observational methods.

%___________________Table of RS Pup distances
\begin{table}
\caption{Observationally determined distances of RS\,Pup. ``IRSB" stands for ``infrared surface brightness".}
\label{distanceobs_rspup}
\begin{tabular}{lll}
\hline
Dist. [pc] & Method & Source \\
\hline
\noalign{\smallskip}
$1\,780 \pm200$ & Light echoes & Havlen (\cite{havlen72a}) \\
\noalign{\smallskip}
2\,040: & Hipparcos parallax & ESA (\cite{esa97}) \\
\noalign{\smallskip}
$2\,111^{+75}_{-73}$ & IRSB & Fouqu\'e et~al. (\cite{fouque03}) \\
\noalign{\smallskip}
$2\,052^{+61}_{-59}$ & IRSB & Storm et~al. (\cite{storm04}) \\
\noalign{\smallskip}
$2\,004\pm 59$ & IRSB (least squares) & Barnes et~al. (\cite{barnes05}) \\
\noalign{\smallskip}
$1\,984\pm 122$ & IRSB (Bayesian) & Barnes et~al. (\cite{barnes05}) \\
\noalign{\smallskip}
$521^{+266}_{-132}$ & Hipparcos re-reduction & van Leeuwen (\cite{vanleeuwen07a}) \\
\noalign{\smallskip}
$694^{+381}_{-181}$ & Hipparcos re-reduction & van Leeuwen et~al. (\cite{vanleeuwen07c}) \\
\noalign{\smallskip}
$1\,830^{+109}_{-94}$ & IRSB & Fouqu\'e et~al. (\cite{fouque07}) \\
\noalign{\smallskip}
\hline
\noalign{\smallskip}
$1\,992\pm 28$ & Light echoes & Present work \\
\noalign{\smallskip}
\hline
\end{tabular}
\end{table}

\subsection{Stellar association}

RS\,Puppis was found to be a likely member of the Pup\,III OB~association discovered by Westerlund~(\cite{westerlund63}). Based on the color-magnitude diagram of about two dozen early-type stars in the region of RS\,Pup, Westerlund derived a distance of 1\,740\,pc and an age of 4 million years for this stellar aggregate. Though this distance implies a spatial coincidence of RS\,Pup with Pup\,III, the age of the Cepheid clearly differs from that of the OB association.

The age of a Cepheid can be reliably estimated from the period-age relationship, whose existence is a consequence of the period-luminosity and mass-luminosity relationships, even if the actual crossing number of the Cepheid through the instability strip is uncertain. The up-to-date form of the period-age relationship is (Bono et al.~\cite{bono05}):
$$\log t = (-0.67 \pm 0.01) \log P + (8.31 \pm 0.08)$$
where $t$ is the age in years. This formula,
valid for the average metallicity of our Galaxy, gives an age of about 17 million years for RS\,Pup excluding the possibility of physical membership of the Cepheid in a 4 million year old stellar grouping. The contradiction between the ages of the Cepheid and the association has already been noted by Havlen~(\cite{havlen72b}), though the age of the Cepheid was much underestimated using the period-age relationship available to him in the early 1970s.

Later studies even cast doubt on the existence of the Pup\,III association (Eggen~\cite{eggen77}). In the meantime Herbst~(\cite{herbst75}) pointed out the existence of an R association (i.e. an association embedded in reflection nebulosity) around RS\,Pup. This Pup\,R3 association was found to be at a distance of $1\,600 \pm 200$\,pc from us, possibly indicating a physical relation between the alleged Pup\,III and Pup\,R3 associations. It is problematic, however, that this R association was defined on only 3 arbitrarily selected stars whose distance moduli happened to be in accordance with that of RS\,Pup obtained by Havlen~(\cite{havlen72a}).

The exhaustive list of the OB associations also includes Pup\,OB3
but a distance of 2\,490\,pc is assigned to it (Mel'nik \& Efremov~\cite{melnik95}).
The source of this distance has not been given.
In a more recent study, Bhatt et~al.~(\cite{bhatt98}) analyzed the polarization
measurements in the direction of Pup\,OB3. Their conclusion was that the
association members can be distinguished from the field stars based on their
polarization characteristics. Bhatt et~al.~(\cite{bhatt98}) did not attempt
to derive a new distance for Pup\,OB3, instead they used
Westerlund's~(\cite{westerlund63}) value of about 1\,700\,pc.

These contradictory statements and especially the age criterion lead us to be
cautious, therefore we refrain from using the cluster/association method for
deriving the distance of RS\,Pup.

%__________________________________Conclusion
\section{Conclusion}

We measured a geometric distance of $d = 1\,992 \pm 28$\,pc to the long-period Galactic Cepheid RS\,Pup using the propagation of light echoes in its circumstellar nebula. Considering the intensity-mean magnitudes $m_B  = 8.446$ and $m_V = 7.009$ from Berdnikov~(private communication), an extinction of $E(B-V) = 0.457 \pm 0.009$ (Fouqu\'e et al.~\cite{fouque07}), and the reddening laws from Laney \& Stobie~(\cite{laney93}), the absolute magnitudes of RS\,Pup in $B$ and $V$ are $M_B = -5.11 \pm 0.05$ and $M_V = -6.09 \pm 0.05$. Our distance is in good agreement with several other estimates, including the value derived by Havlen~(\cite{havlen72a}) of $1\,780 \pm 200$\,pc using the same method. This is, to date, the most accurate geometric distance to a Cepheid.
%and one of the most accurate to any star.
%
A similarly straightforward distance determination using both geometric and physical methods is possible in the far-infrared spectral region. The infrared echo is stimulated by the variable heating of the dust grains caused by the periodic temperature variations during the pulsation of RS\,Pup. The advantage of the infrared band is that the effect of the interstellar extinction can be circumvented and the method is independent of the scattering function of the dust grains. Though this method has been elaborated
by Mayes et~al.~(\cite{mayes85}), it has never been applied until now.

The light echoes of RS\,Pup are arguably the most spectacular example of this fascinating phenomenon. In contrast with supernovae or stellar outbursts (see e.g. Tylenda~\cite{tylenda04}), they are permanently observable using even moderately large telescopes, thanks to the large angular extension of the nebula ($\approx 2.5$\,arcmin). In our images, we can monitor at all times the propagation of more than seven echoes, as visible on the interpolated movie available from the A\&A web site. This nebula undoubtedly holds a wealth of information about the mass-loss history of this star. It will thus be instrumental to understand the evolution of Cepheids. Less than five years away from the centenary of the discovery of the Period--Luminosity relation by Leavitt \& Pickering~(\cite{leavitt12}), RS\,Pup could well become a ``Rosetta stone" for this important class of stars.

%__________________________________Acknowledgements
\begin{acknowledgements}
LSz was supported by the Hungarian OTKA grant T046207. We thank L. N. Berdnikov for providing
us improved intensity-mean magnitudes of RS\,Pup in the $B$ and $V$ bands.
This research made use of the SIMBAD and VIZIER databases at the CDS, Strasbourg (France),
and NASA's Astrophysics Data System Bibliographic Services.
\end{acknowledgements}

%__________________________________Bibliography
{}

\end{document}